\documentclass[final,5p,times,twocolumn]{elsarticle}
\usepackage{graphicx}
\usepackage{amssymb}

\journal{Astroparticle Physics}

\begin{document}

\begin{frontmatter}

\title{Propagation of Ultra-High-Energy Cosmic Ray Nuclei in Cosmic Magnetic Fields and Implications for Anisotropy Measurements}

\author[mpp]{Hajime Takami}
\ead{takami@mpp.mpg.de}
\author[icrr]{Susumu Inoue}
\ead{sinoue@icrr.u-tokyo.ac.jp}
\author[konan]{Tokonatsu Yamamoto}
\ead{tokonatu@konan-u.ac.jp}

\address[mpp]{Max Planck Institute for Physics, F$\ddot{o}$hringer Ring 6, 80805 Munich, Germany}
\address[icrr]{Institute for Cosmic Ray Research, the University of Tokyo, 5-1-5, Kashiwanoha, Kashiwa, Chiba 277-8582, Japan}
\address[konan]{Department of Physics, Konan University, 8-9-1, Okamoto, Higashinada-ku, Kobe, Hyogo 658-8501, Japan}

\begin{abstract}
Recent results from the Pierre Auger Observatory (PAO) indicate that the composition of ultra-high-energy cosmic rays (UHECRs) with energies above $10^{19}$ eV may be dominated by heavy nuclei. An important question is whether the distribution of arrival directions for such UHECR nuclei can exhibit observable anisotropy or positional correlations with their astrophysical source objects despite the expected strong deflections by intervening magnetic fields. For this purpose, we have simulated the propagation of UHECR nuclei including models for both the extragalactic magnetic field (EGMF) and the Galactic magnetic field (GMF). We find that the GMF is particularly crucial for suppressing the anisotropy as well as source correlations. Assuming that only iron nuclei are injected steadily from sources with equal luminosity and spatially distributed according to the observed large scale structure in the local Universe, at the number of events published by the PAO so far (69 events above $5.5 \times 10^{19}$ eV), the arrival distribution of UHECRs would be consistent with no auto-correlation at 95\% confidence if the mean number density of UHECR sources $n_s \gtrsim 10^{-6}$ Mpc$^{-3}$, and consistent with no cross-correlation with sources within 95\% errors for $n_s \gtrsim 10^{-5}$ Mpc$^{-3}$. On the other hand, with 1000 events above $5.5 \times 10^{19}$ eV in the whole sky, next generation experiments can reveal auto-correlation with more than 99\% probability even for $n_s \lesssim 10^{-3}$ Mpc$^{-3}$, and cross-correlation with sources with more than 99\% probability for $n_s \lesssim 10^{-4}$ Mpc$^{-3}$. In addition, we find that the contribution of Centaurus A is required to reproduce the currently observed UHECR excess in the Centaurus region. Secondary protons generated by photodisintegration of primary heavy nuclei during propagation play a crucial role in all cases, and the resulting anisotropy at small angular scales should provide a strong hint of the source location if the maximum energies of the heavy nuclei are sufficiently high.
\end{abstract}


\end{frontmatter}

\section{Introduction} \label{introduction}

The origin of ultra-high-energy cosmic rays (UHECRs) with energies $\gtrsim 10^{19}$ eV is an intriguing mystery in modern astrophysics. Their sources are generally believed to be extragalactic objects, although some Galactic objects may also be viable. In either case, the maximum energy of $10^{20}$ eV can only be achieved in extreme environments \cite{Hillas1984ARAA22p425}. Prominent source candidates suggested so far include active galactic nuclei (AGN) \cite{Biermann1987ApJ322p643,Takahara1990PTP83p1071,Rachen1993AA272p161,Norman1995ApJ454p60,Berezhko2008ApJ684L69,Pe'er2009PRD80p123018,Dermer2010ApJ724p1366,Takami2011APh34p749,Murase2011arXiv1107.5576}, gamma-ray bursts (GRBs) \cite{Waxman1995PRL75p386,Vietri1995ApJ453p883,Murase2006ApJ651L5,Murase2008PRD78p023005}, neutron stars or magnetars \cite{Blasi2000ApJ533L123,Arons2003ApJ589p871,Murase2009PRD79p103001,Kotera2011PhRvD84p023002} and clusters of galaxies \cite{Kang1996ApJ456p422,Inoue2005ApJ628L9,Inoue2007astro-ph0701167}. If UHECRs with energies above $\sim 6 \times 10^{19}$ eV are mainly protons, their propagation distance should be limited by interactions with cosmic microwave background (CMB) photons \cite{Greisen1966PRL16p748,Zatsepin1966JETP4L78,Stecker1969PR180p1264,Berezinsky1971SJNP13p453,Puget1976ApJ205p638}, the so-called Greisen-Zatsepin-Kuz'min (GZK) mechanism, so that they are observable only from sources sufficiently nearby (typically $\lesssim 200$ Mpc). Since all known astrophysical objects are distributed inhomogeneously at such distances in the local Universe, anisotropies in the distribution of UHECR arrival directions are expected as long as cosmic magnetic fields are weak enough to allow quasi-rectilinear propagation of the UHECRs. Searches for correlations between the arrival directions and the celestial positions of potential source candidates should then provide valuable clues to reveal their origin \cite{Waxman1996ApJ483p1,Yoshiguchi:2003vs,Takami2008ApJ678p606}.

The Pierre Auger Observatory (PAO) operating in the southern hemisphere has reported possible evidence of a gradual increase in the average mass of UHECR particles with energies above $10^{18.5}$ eV by analyzing the average depth of the shower maximum $< X_{\rm max} >$ and the root mean square of the shower-to-shower fluctuations of $X_{\rm max}$ ($RMS(X_{\rm max})$), assuming that current hadronic interaction models are realistic at these energies \cite{Abraham2010PRL104p091101}. This is in contrast to measurements of similar quantities for the northern sky by the High Resolution Fly's Eye (HiRes) experiment that are consistent with an UHECR composition dominated by protons \cite{Abbasi2005ApJ622p910,Abbasi2010PRL104p161101}. If UHECRs consisted mainly of heavy nuclei rather than protons as indicated by the PAO results, their deflections during propagation in magnetized extragalactic and Galactic environments would be much larger than those for protons, and the hopes for observing anisotropy may be considerably weakened. On the other hand, PAO has seen statistically significant correlations between the arrival directions of UHECRs with energies above $\sim 6 \times 10^{19}$ eV and the projected positions of AGN with $z \leq 0.018$ within an angular scale of $3.1^{\circ}$ \cite{Abraham2007Sci318p938,Abraham2007Aph29p188} \footnote{Contrastingly, HiRes has not found significant cross correlations with nearby extragalactic objects based on the same analysis method as PAO \cite{Abbasi2008Aph30p175}. However, the two results cannot be naively compared because the exposure of HiRes is smaller than that of PAO, and they also correspond to different regions of the sky.}. This does not necessarily point to AGN as the true sources, since they may only be tracers of the more numerous population of galaxies that make up the anisotropic, local cosmic matter distribution, and there are also considerable uncertainties in the angular displacement due to deflections by both the extragalactic magnetic field (EGMF) and the Galactic magnetic field (GMF). Nevertheless, this does imply that UHECR sources are associated with the matter distribution. Independent analyses of the PAO data have confirmed the correlations of UHECR arrival directions with various kinds of nearby astrophysical objects \cite{Kashti2008JCAP05p006,George2008MNRAS388L59,Ghisellini2008MNRAS390L88,Takami2008JCAP06p031}. Such correlations were predicted in several theoretical studies under the assumption of protons as the primary UHECR particles \cite{Waxman1996ApJ483p1,Yoshiguchi:2003vs,Takami2008ApJ678p606}. We must beware that at the moment, the positional correlations (or anisotropy) are observed only {\it above} $\sim 6 \times 10^{19}$ eV, whereas the PAO results on the composition are available only {\it up to} $\sim 6 \times 10^{19}$ eV due to the lack of event statistics. We cannot rule out the possibility that protons become dominant above this energy and produce the anisotropy. However, there are presently no hints of such a sudden change in the composition, and the simplest extrapolation of the measured trends of $< X_{\rm max} >$ and $RMS(X_{\rm max})$ to higher energies would entail a composition dominated by heavy nuclei. Since the appearance of anisotropies or cross correlations with astrophysical objects in the case of UHECR nuclei is not trivial, deeper, relevant studies of UHECR propagation are warranted in order to achieve a consistent interpretation of the current data set as well as to quantify the prospects for future observations.

Propagation of UHE nuclei has been studied previously by many authors from various perspectives \cite{Stecker1969PR180p1264,Berezinsky1971SJNP13p453,Puget1976ApJ205p638,Anchordoqui1998PRD57p7103,Epele1998JHEP10p009,Stecker1999ApJ512p521,Bertone2002PRD66p103003,Yamamoto2004APh20p405,Hooper2005APh23p11,Ave2005APh23p19,Armengaud2005PRD72043009,Sigl2005JCAP10p016,Allard2005AA443L29,Khan2005APh23p191,Harari2006JCAP11p012,Allard2007APh27p61,Hooper2007APh27p199,Anchordoqui2007PhRvD76p123008,Arisaka2007JCAP12p002,Globus2008AA479p97,Hooper2008PRD77p103007,Aloisio2008arXiv08024452,Allard2008JCAP10p033,Hooper2010APh33p151,Aloisio2010arXiv10062484}. Many have studied the photodisintegration process and consequent spectra in detail without accounting for magnetic fields, but only a few have focused on the effects of the magnetic fields. Ref. \cite{Yamamoto2004APh20p405} calculated the propagation of UHE nuclei in a uniform turbulent EGMF model and examined their trajectories and photodisintegration interactions in intergalactic photon fields, concluding that a characteristic feature in the spectrum may result. Refs. \cite{Armengaud2005PRD72043009,Sigl2005JCAP10p016} discussed the propagation of nuclei in a structured EGMF model obtained through a numerical simulation of cosmological structure formation, showing that while heavy nuclei can be strongly deflected ($>20^{\circ}$) by the EGMF, nuclei (and their secondaries) with relatively small deflections can generate anisotropy on intermediate angular scales.

In this paper, we study the propagation of UHECR nuclei in cosmic radiation and magnetic fields under different assumptions for the source properties and the fields, focusing on the resulting distribution of arrival directions. In addition to a detailed treatment of the photopair, photomeson and photodisintegration interactions of UHE nuclei with background photons, deflections in GMF and EGMF are taken into account for both primary nuclei as well as secondary nuclei arising from photodisintegration. By statistically analyzing the calculated arrival distributions, we discuss the implications for current and future observations of anisotropy and cross correlations with sources.

Motivated by the recent PAO result, we adopt a model for the UHECR source distribution that follows the density of large-scale structure actually observed in the local Universe. This is in contrast to Refs. \cite{Armengaud2005PRD72043009,Sigl2005JCAP10p016} who took the sources to be related to the matter distribution in a numerical simulation, but not reflecting the actual Universe. Because photodisintegration limits the propagation distance of heavy nuclei in a way similar to the GZK mechanism for protons, a realistic account of local source inhomogeneity is essential. Unlike most previous work on the subject, here we also account for the GMF as well as the EGMF, since the former unavoidably affects all UHECRs arriving at Earth. Concerning the UHECR source composition, a wide variety of possibilities are currently allowed that are consistent with both the observed spectrum and the composition indicators $X_{\rm max}$ and $RMS(X_{\rm max})$ (e.g., \cite{Hooper2010APh33p151}). This study assumes a pure iron composition, which can also reproduce the highest-energy spectrum, and which would be the most pessimistic situation for producing anisotropy or source correlations in view of their large magnetic deflections.

This paper is laid out as follows. In Section \ref{method} we describe our models and methods of calculation and statistical analysis. Parametrization of photodisintegration cross sections to calculate the mean free paths of nuclei in cosmic background radiation fields is summarized in Section \ref{method_pd}. Section \ref{method_mf} is dedicated to our models for the GMF and EGMF. A method of calculation for nuclear propagation is introduced in Section \ref{method_prop}, and that for UHECR arrival distribution is presented in Section \ref{method_ad}. The contents of these two subsections are essentially based on Ref. \cite{Yoshiguchi2003ApJ586p1211} that dealt with UHE proton propagation in intergalactic space, and here we extend their methods to heavy nuclei. We describe our statistical methods for studying anisotropy and cross correlations with sources in Section \ref{method_stat}. In Section \ref{results}, we present our simulation results, compare with the current data \cite{Abraham2007Aph29p188,Abreu2010APh34p314}, and discuss the implications thereof for the sources of UHE nuclei. We also give special attention to Centaurus (Cen) A, for which several PAO events positionally correlate \cite{Abreu2010APh34p314}. We summarize in Section \ref{summary}. 

\section{Details of Calculations} \label{method}

\subsection{Photodisintegration} \label{method_pd}

A nucleus emits its constituent nucleons by interactions with photons ({\it photodisintegration}). Since the number of emitted nucleons per interaction is stochastically determined following branching ratios dependent on interaction energy, a lot of disintegration paths are realized during propagation in photon fields. However, all the branches in photodisintegration are not well known yet. This study adopts a simple disintegration track on the nuclear chart proposed by Ref. \cite{Puget1976ApJ205p638}. There is only one nuclear species for each nuclear mass number $A$ in this treatment.

\begin{figure}[t]
\begin{center}
\includegraphics[width=0.95\linewidth]{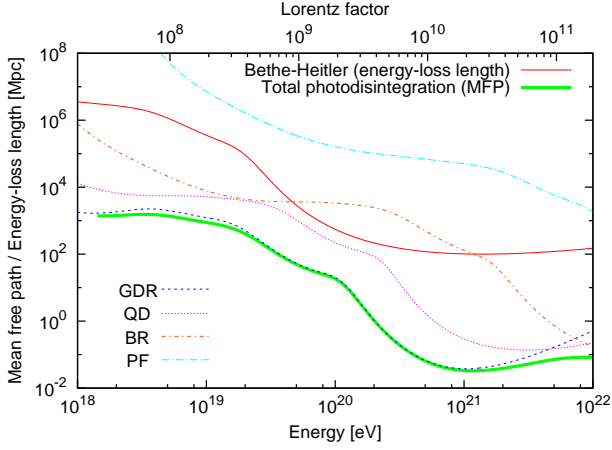}
\caption{Energy-loss length of Bethe-Heitler pair creation ({\it red}) and mean free path of total photodisintegration ({\it green}) in a model of cosmic background radiation field \cite{Kneiske2004AA413p807} for $^{56}_{26}$Fe. The contribution of individual processes of photodisintegration is also shown for comparison.} 
\label{fig:mfp}
\end{center}
\end{figure}

Photodisintegration can be phenomenologically classified into 4 processes: giant dipole resonance (GDR), quasi-deuteron (QD) process, baryonic resonance (BR), and photofragmentation (PF). Since the importance of the latter two processes appears above $10^{22}$ eV for Fe nuclei in the CMB because of their very high threshold energy (fig. \ref{fig:mfp}, and see also Section \ref{method_prop}), our assumption of $E_{\rm max}^{\rm Fe} = 10^{21.5}$ eV allows to neglect these processes. This study adopts the parametrization of photodisintegration cross-sections developed by Ref. \cite{Rachen1996phd}. The mean free paths of nuclear photodisintegration in cosmic background photon fields can be calculated from the cross-sections (explained in Section \ref{method_prop}).

GDR has the lowest threshold energy in the photodisintegration processes, typically $\sim 10$ MeV in the nucleus rest frame.  A nucleus emits several nucleons or $\alpha$ particles, but one nucleon emission is dominated. The treatment of Ref. \cite{Rachen1996phd} approximated GDR as a process emitting one nucleon. The cross-section of GDR is parametrized by a Lorenzian function 
\begin{equation}
\bar{\sigma}_L(\epsilon',~E_0,~\Gamma) = \frac{\Gamma^2}{(\epsilon' - E_0)^2 + \Gamma^2}, 
\end{equation}
as 
\begin{equation}
\sigma_{A,{\rm GDR}}(\epsilon') = \hat{\sigma}_0 ~A^{7/6} ~\bar{\sigma}_{\rm L}(\epsilon',~\hat{\epsilon}'_{\rm GDR} A^{-1/6},~\hat{\Gamma}_{\rm GDR} A^{-1/6}), 
\end{equation}
for $\epsilon' \geq \epsilon'_{A,{\rm th,GDR}}$ and 0 otherwise. Here $\epsilon'$ is photon energy in the nuclear rest frame, $\hat{\sigma}_0 = 0.72$ mbarn, $\hat{\epsilon'}_{\rm GDR} = 35.3$ MeV, and $\hat{\Gamma}_{\rm GDR} = 15.1$ MeV. The threshold energy of this process is also scaled as a function of $A$, $\epsilon'_{A,{\rm th,GDR}} = 16.6~A^{-1/6}$ MeV.

The quasi-deuteron process is treated as a two nucleon emission process. The parametrization of its cross-section is 
\begin{equation}
\sigma_{A,{\rm QD}}(\epsilon') = 0.55~A^{5/4}~\frac{(\epsilon'/\epsilon'_{A,{\rm th,QD}} - 1)^{3/2}}{(\epsilon'/\epsilon'_{A,{\rm th,QD}})^3} ~~{\rm mbarn}
\end{equation}
for $\epsilon' \geq \epsilon'_{A,{\rm th,QD}}$ where $\epsilon'_{A,{\rm th,QD}} = 33.3~A^{-1/6}$ MeV.

While the scaling laws on $A$ can well reproduce experimental data for large $A$ nuclei, the behavior of small nuclei ($A < 10$) is different because of their small number of nucleons. Thus, other analytical functions of the cross-sections are used for deuteron, trinucleon (triton or $^3{\rm He}$), helium, and $^9_4$Be instead of $\sigma_{A,{\rm GDR}}(\epsilon')$, $\sigma_{A,{\rm QD}}(\epsilon')$. $^9_4$Be is assumed to decay into two $^4_2$He and one nucleon. We also assume that a nucleus with $A = 8$ which is produced by QD of a particle with $A = 10$ immediately decay into two $^4_2$He because there is no stable particle with $A = 8$. Therefore, nuclei between $A = 5$ and $A = 8$ are not observed at the Earth in this treatment.

The photoabsorption cross-section of a deuteron was derived by Ref. \cite{Bethe1950PR77p647} as
\begin{equation}
\sigma_d(\epsilon') = \frac{\sigma_{\rm BP}(\epsilon',~B_d)}{1 - a r_{\rm eff}}; 
~~~a = \frac{\sqrt{m_p B_d}}{\hbar}, 
\end{equation}
where 
\begin{equation}
\sigma_{\rm BP} (\epsilon',~B) = \frac{\sigma_{\rm Tp}}{\alpha} 
\frac{m_p c^2}{B} \frac{(x - 1)^{3/2}}{x^3}; ~~~x = \frac{\epsilon'}{B}, 
\end{equation}
is a photoabsorption cross-section for a zero range nuclear force \cite{Bethe1935PRSA148p146}. $\hbar$, $m_p$, $c$, $\alpha$, $B_d = 2.227$ MeV, $r_{\rm eff}=1.79$ fm, and $\sigma_{Tp} = 1.97 \times 10^{-4}$ mbarn are the Planck constant, the proton mass, the speed of light, the fine structure constant, the nuclear binding energy of a deuteron, the effective range of nuclear force, and the cross-section of Thomson scattering for proton mass, respectively. After this reaction, a deuteron is broken to a proton and neutron.

A trinucleon has two interaction channels; it emits a nucleon and is changed into a deuteron, or three body decay. The cross-sections of these processes can be well fitted by using $\sigma_{\rm BP}$, 
\begin{eqnarray}
\sigma^{\rm T}_1 (\epsilon') &=& 1.4 ~\sigma_{\rm BP}(\epsilon',~5.8 {\rm MeV}) 
~~~ {\rm for} ~ {\rm T}_3(\gamma,N){\rm D}_2 \\
\sigma^{\rm T}_2 (\epsilon') &=& 1.7 ~\sigma_{\rm BP}(\epsilon',~7.3 {\rm MeV}) 
~~~ {\rm for} ~ {\rm T}_3(\gamma,3N)
\end{eqnarray}

The cross-sections of one or two particle emission of $\alpha$-particle are parametrized as 
\begin{eqnarray}
\sigma^{\alpha}_1 (\epsilon') &=& 
(3.6 {\rm mbarn}) ~{\rm Pl}(\epsilon',~19.8 {\rm MeV},~27 {\rm MeV},~5) \\
\sigma^{\alpha}_2 (\epsilon') &=& 1.4~\sigma_{\rm BP}(\epsilon',~26.1 {\rm MeV}), 
\end{eqnarray}
respectively, where 
\begin{equation}
{\rm Pl}(x; x_{\rm th}, x_{\rm max}, \alpha) = 
\left( \frac{x - x_{\rm th}}{x_{\rm max} - x_{\rm th}} \right)^{{\tilde \alpha} - \alpha} 
\left( \frac{x}{x_{\rm max}} \right)^{-{\tilde \alpha}} 
\Theta \left( x - x_{\rm th} \right). 
\end{equation}
Here, $\Theta (x)$ is a step function and ${\tilde \alpha} = \alpha \left( x_{\rm max} / x_{\rm th} \right)$.

The cross-section of photo-absorption of $^9_4$Be is parametrized as 
\begin{equation}
\sigma^{^9{\rm Be}} (\epsilon') = \sigma^{^9{\rm Be}}_1 (\epsilon') + \sigma^{^9{\rm Be}}_2 (\epsilon'),
\end{equation}
where 
\begin{eqnarray}
\sigma^{^9{\rm Be}}_1 (\epsilon') &=& 
1.5 ~\bar{\sigma}_L(\epsilon', 1.7 {\rm MeV}, 0.3 {\rm MeV}) \nonumber \\
&& ~ + 1.6 ~\bar{\sigma}_L(\epsilon', 10.0 {\rm MeV}, 10.0 {\rm MeV}) \nonumber \\
&& ~ + 3.5 ~\bar{\sigma}_L(\epsilon', 25 {\rm MeV}, 15 {\rm MeV}) 
~~{\rm mbarn} \\
\sigma^{^9{\rm Be}}_2 (\epsilon') &=& 9 ~\sigma_{\rm BP}(\epsilon',16.9 {\rm MeV}).
\end{eqnarray}

We should also treat the propagation of protons as secondaries. For protons, photopion production plays an important role above $6 \times 10^{19}$ eV. The cross-section and energy-loss rate of protons are calculated by a photomeson production event generator {\it SOPHIA} \cite{Mucke1999CPC124p290}.

\subsection{EGMF and GMF} \label{method_mf}
 
Based on the cosmological structure formation theory, density fluctuations produced by inflation have gravitationally grown and form large-scale structure of the current universe. Galaxy surveys have confirmed through galaxy distribution that matter distribution is highly structured. Since magnetic fields has been also amplified through the compression and dynamo mechanism of astrophysical plasmas simultaneously, they could be also structured following the matter distribution. Several simulations of cosmological structure formation with magnetic fields have confirmed this expectation qualitatively \cite{Sigl2003PRD68p043002,Sigl2004PRD70p043007,Dolag2005JCAP01p009,Ryu2008Science320p909}, but EGMF structures highly depend on modeling due to few observational constraints. The difference appears, for example, in the volume filling factor of strong EGMFs (see Ref. \cite{Kotera2011ARA&A49p119}). Therefore, a variety of EGMF modelings gives us different results on the deflection of UHECRs and source identification possibility even for protons \cite{Sigl2003PRD68p043002,Sigl2004PRD70p043007,Dolag2005JCAP01p009,Takami2006ApJ639p803,Das2008ApJ682p29,Kotera2008PRD77p123003}. In addition to the uncertainty, it is technically difficult to calculate the arrival distribution of UHECRs with a complex EGMF structure because only a very small fraction of UHECRs injected from sources can arrive at the Earth. Ref. \cite{Takami2006ApJ639p803} developed a numerical method to solve this problem for UHE protons, but this method cannot apply to the cases of heavy nuclei owing to the stochastic nature of photodisintegration. In order to avoid the uncertainty of EGMF modeling and the numerical difficulty, this study assumes a uniform turbulent EGMF model with the strength of $B_{\rm EGMF}$, the correlation length of $\lambda_{\rm EGMF}$, and the Kolmogorov power spectrum \cite{Yoshiguchi2003ApJ586p1211}. The averaged property of EGMFs has been constrained by Faraday rotation measurements of distant quasars as ${B_{\rm EGMF}} {\lambda_{\rm EGMF}}^{1/2} < (10 {\rm nG}) (1 {\rm Mpc})^{1/2}$ \cite{Ryu1998AA335p19,Blasi1999ApJ514L79}. A recent analysis of CMB and matter power spectra derives an upper limit of the primordial magnetic field at present, which is regarded as a magnetic field in voids, of $B_{\rm v} < 3$ nG (95 \% C.L.) for the coherent length of $\lambda_{\rm v} = 1$ Mpc \cite{Yamazaki2010PRD81p023008}. Several simulations of cosmological structure formation with magnetic fields indicate $B_{\rm v} \ll 1$ nG \cite{Dolag2005JCAP01p009,Das2008ApJ682p29}. On the other hand, pair-halo and/or pair-echo emission from TeV blazars may provide lower bounds, suggesting $B_{\rm v} \gtrsim 10^{-18}$ - $10^{-17}$ G from the data of {\it Fermi} Large Area Telescope \cite{Dolag2011ApJ727L4,Dermer2011ApJ733L21,Takahashi2012ApJ744L7}. Even magnetic fields in voids are negligibly weak, the effective magnetic fields of cosmic structures such as filaments and clusters of galaxies may have $B_{\rm EGMF} {\lambda_{\rm EGMF}}^{1/2} \lesssim 1$ nG Mpc$^{1/2}$ \cite{Takami2011arXiv1110.3245}. In this study $B_{\rm EGMF}$ is treated as a parameter, while $\lambda_{\rm EGMF} = 1$ Mpc is fixed which is motivated by the average separation distance between galaxies.

The GMF has been relatively well observed compared to EGMFs through Faraday rotation measurements of radio pulsars and extragalactic radio sources. It has a regular component with a few $\mu$G on average in the Galactic disk, whose shape is similar to the spiral structure of the Milky Way. There are also turbulent fields with the magnetic strength 0.5-2 times as large as that of the spiral component \cite{Beck2000SSR99p243}. Its correlation length has been lower-limited to $\lambda_{\rm GMF} \sim 100$ pc due to the finite resolution of radio observations. Both the components significantly contribute to the total deflections of UHE nuclei \cite{Giacinti2011APh35p192}. However, at present, even the newest data are difficult to discriminate the regular structure between the bisymmetric and axisymmetric models \cite{Pshirkov2011ApJ738p192}, and the difference produces significantly different propagation of UHECRs especially in the case of heavy nuclei, e.g., nuclear propagation is sensitive to GMF models \cite{Takami2010ApJ724p1456}. In order to avoid the model dependence, we simply assume that the deflection angles of UHECRs do not depend on directions. In Ref. \cite{Giacinti2010JCAP08p036}, a GMF model with a weak dipole field predicted a typical deflection angle of $40^{\circ}$-$50^{\circ}$ for irons with $\sim 6 \times 10^{19}$ eV on average in the whole sky. Assuming a rigidity scaling additionally, we set the typical deflection angles of UHECRs in GMF to be $\theta_{\rm GMF} = 1.0^{\circ} Z ( E / 10^{20} {\rm eV} )^{-1}$ where $E$ is the energy of UHECRs.

\subsection{Calculation of Propagation} \label{method_prop}

The calculation of nuclear propagation can be divided into two parts; propagation in extragalactic space and in Galactic space. The size of the Galactic space (assumed to be 40 kpc) is much smaller than the energy-loss length of Bethe-Heitler pair creation. On the other hand, the mean free path of photodisintegration is close to the size of Galactic space above $10^{20.5}$ eV for $^{56}_{26}$Fe. However, it does not affect the propagation in the Galaxy because particles traveled for longer distances than the mean free path in extragalactic space before entering the Milky Way. Thus, the nuclear reactions can be neglected during the propagation in the Galaxy. This allows us to treat the effect of the propagation as modifications of the arrival directions of UHECRs. The modifications will be described in Section \ref{method_ad}. Below, we introduce a method of the propagation of UHECRs in extragalactic space.

We calculate the propagation trajectories of UHECRs including secondary particles in extragalactic space, and estimate the nuclear species, energies, deflection angles of nuclei arriving at Earth. In our method, the source of primary nuclei ($^{56}_{26}$Fe) is located at the center of coordinates in extragalactic space. The primary nuclei are ejected from the source isotropically, and then their trajectories are calculated step by step including secondary particles taking magnetic deflections, photodisintegration, and Bethe-Heitler pair creation into account. Adiabatic energy-loss due to cosmic expansion is neglected because the energy-loss rate is much lower than the others. Since this study focuses on the highest energy range, $E > 5.5 \times 10^{19}$ eV, the calculation of the trajectories is finished at 500 Mpc from the source. We confirmed that nuclei with the energy above $E > 5.5 \times 10^{19}$ eV cannot arrive at the Earth from more distant sources even for rectilinear propagation. The nuclear species, energy, and deflection angles of the particles are recorded every 1 Mpc. We divide a logarithmic energy bin into 10 per decade, i.e., $10^{19.75}$-$10^{19.85}$ eV, $10^{19.85}$-$10^{19.95}$ eV, $\cdots$, $10^{21.45}$-$10^{21.55}$ eV, and inject 10,000 Fe nuclei in each bin. The record allows us to make the distribution of the nuclear species, energy and deflection angles of the particles at 500 distances. For instance, when we consider UHECRs injected from a source with the distance of 100 Mpc, the 100th distributions of them are regarded as their distributions at the Earth.

In the calculation of propagation of nuclei, photo-disintegration processes are treated as a stochastic process, i.e., by a Monte-Carlo method. The mean free path of photo-disintegration process $i$ ($=$ GDR, QD) of a nucleus with the nuclear mass number $A$ and the energy $E$ can be calculated by using the expressions of cross-sections described in Section \ref{method_pd} as 
\cite{Protheroe1996Aph4p253}, 
\begin{equation}
\frac{1}{\lambda_{A,i}(E)} = \frac{1}{8 \beta E^2} 
\int_{\epsilon_{\rm th}}^{\infty} \frac{d\epsilon}{\epsilon^2} 
\frac{dn_{\gamma}}{d\epsilon}(\epsilon) 
\int_{s_{\rm min}}^{\rm s_{\rm max}} ds \sigma_{A,i}(s) 
\left( s - {M_A}^2 c^4 \right), 
\label{Eq:mfp}
\end{equation}
where $\beta$, $\epsilon$, $s$, and $M_A$ are nuclear velocity normalized by $c$, the energy of photons in the laboratory frame, the Lorentz invariant mass squared, and nucleus mass, respectively. $\sigma_{A,i}(s)$ is cross-section of photo-disintegration process considered, $s_{\rm min} = {M_A}^2 c^4 + 2 M_A c^2 \epsilon'_{A,{\rm th},i}$, $s_{\rm max} = {M_A}^2 c^4 + 2 E \epsilon ( 1 + \beta )$, and $\epsilon_{\rm th} = (s_{\rm min} - {M_A}^2 c^4) \left[ 2E ( 1 + \beta ) \right]^{-1}$. $M_A$ is simply defined as $M_A = A m_n$, where $m_n = 931.494$ MeV is nucleon mass. $dn_{\gamma}/d\epsilon$ is the number density of extragalactic background light (EBL) photons including CMB photons. This study adopts the low-IR model of Ref. \cite{Kneiske2004AA413p807}. Since the Bethe-Heitler pair creation process is treated in a continuous energy-loss approximation (see below), the occurrence of reactions is judged by the method discussed in Ref. \cite{Stanev2000PRD62p093005}. GDR and QD are treated independently. We approximate that the Lorentz factor of nuclei is unchanged by photodisintegration.

For the energy-loss length of the Bethe-Heitler pair creation, we adopt an analytical fitting formula developed by Ref. \cite{Chodorowski1992ApJ400p181}, 
\begin{equation}
- \frac{d\gamma}{dt} = \alpha {r_{\rm e}}^2 c Z^2 \frac{m_e}{M_A} 
\int_2^{\infty} d\kappa \frac{d n}{d\epsilon} 
\left( \frac{\kappa}{2\gamma} \right) 
\frac{\varphi(\kappa)}{\kappa^2}, 
\end{equation}
where $r_{\rm e}$ is the classical electron radius, $m_e$ is electron mass, and $\kappa = \epsilon / m_e c^2$. $\varphi(\kappa)$ is an integral similar to the first integral of Eq. \ref{Eq:mfp}, but is parametrized as 
\begin{eqnarray}
\varphi(\kappa) = \left\{
\begin{array}{ll}
\frac{\pi}{12} (\kappa - 2)^4 
\left[ 1 + \sum_{i=1}^4 c_i (\kappa - 2)^i \right]^{-1} & \kappa \leq 25 \\
\kappa \left[ \sum_{i=0}^3 d_i (\ln \kappa)^i \right] 
\left[ 1 - \sum_{i=1}^3 f_i \kappa^{-i} \right]^{-1} & \kappa > 25, 
\end{array}
\right.
\end{eqnarray}
where $c_1 = 0.8048$, $c_2 = 0.1459$, $c_3 = 1.137 \times 10^{-3}$, $c_4 = -3.879 \times 10^{-6}$, $d_0 = -86.07$, $d_1 = 50.96$, $d_0 = -14.45$, $d_0 = 8/3$, $f_1 = 2.910$, $f_2 = 78.35$, and $f_3 = 1837$. 

\subsection{Calculation of Arrival Distribution} \label{method_ad}

The arrival distribution of UHECRs can be calculated using the distributions obtained by the method written in Section \ref{method_prop} with an assumed source model. 

We adopt a source model developed in Ref. \cite{Takami2006ApJ639p803}, in which UHECR sources are distributed following galaxy distribution actually observed. The source distribution of this model is based on the galaxy catalog of the Infrared Astronomical Satellite (IRAS) Point Source Redshift Survey (PSCz) \cite{Saunders2000MNRAS317p55}. This galaxy catalog covers $\sim 84\%$ of the whole sky, and is enough to construct the model of matter distribution in local Universe (e.g., Ref. \cite{Dolag2005JCAP01p009}). Selection effects of galaxies (e.g., unobserved sky and galaxies fainter than the flux limit) are compensated by a luminosity function \cite{Takeuchi2003ApJ587L89}, and then the model of galaxy distribution following matter distribution is constructed within 200 Mpc. Outside 200 Mpc, we assume isotropic distribution due to the lack of completeness of galaxies in the catalog. This assumption does not change the results of this study, since distant sources do not contribute to anisotropy. We select galaxies randomly from the galaxy distribution following $n_s$, which is the number density of UHECR sources as a parameter, and regard a set of the selected galaxies as a source distribution. Although, $n_s$ has been estimated as $\sim 10^{-4}$ Mpc$^{-3}$ \cite{Takami2009Aph30p306,Cuoco2009ApJ702p825} on the assumption of pure proton composition, it is not clear in the case of heavy nuclei. A smaller $n_s$ may be allowed owing to larger deflections. In this study, we survey over $n_s = 10^{-3}$-$10^{-7}$ Mpc$^{-3}$. Corresponding objects are shown in Table \ref{tab:nd}. All the analyses in this study are limited to steady sources. We assume that all the sources emit UHECRs with the same power for simplicity. The injection spectrum of the particles is assumed to be a power-law $\propto E^{-2.0}$ with $E^{Fe}_{\rm max} = 10^{21.5}$ eV. We confirmed that this spectrum can reproduce observed spectrum above $6 \times 10^{19}$ eV within uncertainties.

In Table \ref{tab:nd}, Fanaroff-Riley (FR) I galaxies and BL Lac objects are in different categories. However, following a unification theory of radio-loud AGNs, these originate from the same population with the only difference due to the viewing angle \cite{Urry1995PASP107p803}. Thus, if the deflection angles of UHECRs during propagation in extragalactic space is much larger than a typical opening angle of their jets ($\sim 0.1$ radian), we can see these objects as the essentially same objects in the viewpoint of UHECR sources.

An assumed $n_s$ constrains the energy budget of UHECR emission per source. The observed spectrum of UHECRs requires $\sim 10^{44}$ erg Mpc$^{-3}$ yr$^{-1}$ \cite{Waxman1995ApJ452L1,Berezinsky2006PRD74p043005,Murase2008ApJ690L14} for UHECR emission above $10^{19}$ eV. Thus, the energetics of UHECR emission required per source is $\sim 3 \times 10^{40} (n_s / 10^{-4} {\rm Mpc}^{-3})^{-1}$ erg s$^{-1}$. 

The flux contribution of each source is inversely proportional to the distance squared. Following the injection spectrum weighted by this factor and the distribution of UHECRs in deflections, energies, and nuclear mass number (see Section \ref{method_prop}), we can simulate the arrival distribution of UHECRs. In addition, we can add the contribution of the GMF in the arrival distribution. Once the arrival directions of UHE nuclei are realized following the method above (considering only EGMF), the arrival directions are modified following the two dimensional Gaussian distribution with the zero mean and the standard deviation of $\theta_{\rm GMF}$. 

\begin{table}
\caption{Local number densities of active objects.}
\begin{center}
\begin{tabular}{lll}
\hline
Objects & Density [Mpc$^{-3}$] & Refs. \\ \hline \hline
Seyfert galaxy & $1 \times 10^{-3}$ & \cite{Ulvestad2001ApJ558p561} \\
Bright quasar & $1 \times 10^{-6}$ & \cite{Hewett1993ApJ406L43} \\
Fanaroff-Riley I & $8 \times 10^{-5}$ & \cite{Padovani1990ApJ356p75} \\
Fanaroff-Riley II & $3 \times 10^{-8}$ & \cite{Woltjer1990agnconf} \\
BL Lac objects & $3 \times 10^{-7}$ & \cite{Woltjer1990agnconf} \\ \hline
\end{tabular}
\label{tab:nd}
\end{center}
\end{table}

\begin{figure*}[t]
\begin{center}
\includegraphics[width=0.48\linewidth]{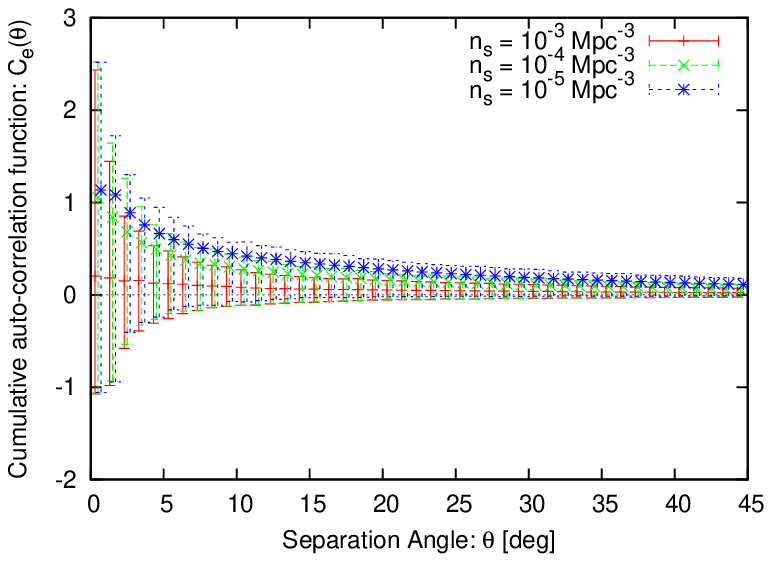} \hfill
\includegraphics[width=0.48\linewidth]{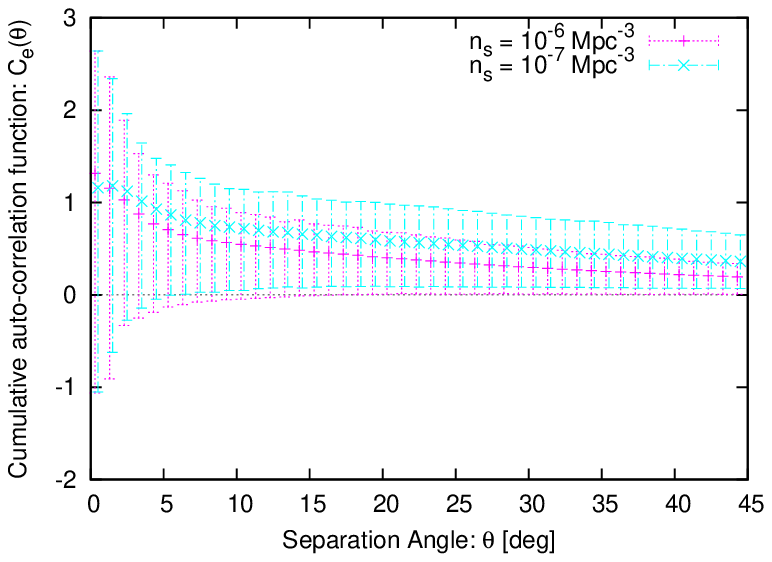}
\caption{Cumulative auto-correlation functions of simulated events for the source models with $n_s = 10^{-3}$, $10^{-4}$, $10^{-5}$ Mpc$^{-3}$ ({\it left}) and $10^{-6}$, $10^{-7}$ Mpc$^{-3}$ ({\it right}). The simulated event sets mimic the recently published events of the PAO, i.e., 69 events with the energy above $5.5 \times 10^{19}$ eV, assuming the PAO aperture. Both EGMF ($B_{\rm EGMF} = 1$ nG) and the GMF are taken into account. Points are the averaged values of the cumulative auto-correlation functions over 1000 source realizations for each $n_s$. The error bars correspond to 68\% (1$\sigma$) errors.} 
\label{fig:acor}
\end{center}
\end{figure*}

\subsection{Statistical Quantities} \label{method_stat}

Anisotropy in the arrival distribution of UHECRs can be quantified by a cumulative auto-correlation function \cite{Takami2011PThPh126p1123}, 
\begin{equation}
C_{\rm e}(\theta) = \frac{EE(<\theta) - 2EE'(<\theta) + E'E'(<\theta)}{E'E'(<\theta)}, 
\label{eq:acor}
\end{equation}
where $E$ and $E'$ denote UHECR events and events randomly distributed following the aperture of a UHECR detector, respectively. $EE(<\theta)$ is the number of UHECR event pairs with the separation angle less than $\theta$ divided by the total number of pairs. $EE'(<\theta)$ and $E'E(<\theta)$ are defined similarly to $EE(<\theta)$. This function is essentially equivalent to the quantity that the number of UHECR event pairs with the separation angle within $\theta$ is divided by the total number of pairs and by a solid angle within $\theta$ if the aperture of UHECR observatories is uniform. However, since, in general, the aperture is not uniform in the cases of ground-based detectors, the non-uniformity should be corrected. $E'$ can correct this in this formula. This function was originally proposed by Ref. \cite{Landy1993ApJ412p64} as a differential auto-correlation function. This correction simplifies the interpretation of the cumulative auto-correlation function; $C_{\rm e}(\theta) > 0$ shows the positive excess of UHECR distribution within $\theta$ compared to isotropic distribution.

Many tests of the correlation between the arrival directions of UHECRs and the positions of their source candidates are based on the excess of the number of events within a circle centered by the source candidates over randomly distributed events. This study also adopts a similar method to study the correlation, using a cumulative cross-correlation function, which is defined similarly to the cumulative auto-correlation function, 
\begin{equation}
C_{\rm es}(\theta) = \frac{ES(<\theta) - ES'(<\theta) - E'S(<\theta) + E'S'(<\theta)}{E'S'(<\theta)}, 
\label{eq:ccor}
\end{equation}
where $S$ and $S'$ denote UHECR sources and sources randomly distributed following selection effects of the catalog of the source candidates. $S'$ is isotropically distributed in the entire sky in this study because of the correction of selection effects (see Section \ref{method_ad}). $ES(<\theta)$ is the normalized number of pairs between UHECR events and sources with the separation angle less than $\theta$. $ES'(<\theta)$, $E'S(<\theta)$, and $E'S'(<\theta)$ are similarly defined. The interpretation of this quantity is also simple; $C_{\rm es}(\theta) > 0$ means positive correlation within $\theta$.

In order to calculate Eqs.\ref{eq:acor} and \ref{eq:ccor}, the distribution of $E'$ should be understood. The aperture of a ground array depends on the declination of observed directions reflecting the daily rotation of the Earth. The declination dependence of the aperture can be analytically estimated as \cite{Sommers:2000us}
\begin{equation}
\omega(\delta) \propto \cos(a_0) \cos(\delta) \sin(\alpha_m) 
+ \alpha_m \sin(a_0) \sin(\delta), 
\end{equation}
where $\alpha_m$ is given by
\begin{eqnarray}
\alpha_m = \left\{
\begin{array}{ll}
0 & {\rm if}~\xi > 1 \\
\pi & {\rm if}~\xi < -1 \\
\cos^{-1}(\xi) & {\rm otherwise}
\end{array}
\right.
\end{eqnarray}
and
\begin{equation}
\xi \equiv \frac{\cos(\Theta) - \sin(a_0) \sin(\delta)}{\cos(a_0) \cos(\delta)},
\end{equation}
when observation time is sufficiently larger than a day. Here, $a_0$ is the celestial latitude of the ground array and $\Theta$ is the zenith angle for a data quality cut because of experimental reasons. For the PAO, $a_0 = -35.2^{\circ}$ and $\Theta = 60^{\circ}$ \cite{Abraham2007Sci318p938,Abraham2007Aph29p188}. The $E'$ sample is generated following this equation. We set the number of $E'$ events to be 200000 in order that the distribution of random events reflects the PAO aperture sufficiently. Also, we checked that this particular choice of the number of random events does not affect results.

\section{Results} \label{results}

\subsection{Anisotropy} \label{aniso}

Fig.\ref{fig:acor} shows the cumulative auto-correlation functions of simulated events with 68\% errors in the cases of $n_s = 10^{-3}$-$10^{-7}$ Mpc$^{-3}$. They are divided into two panels for visibility. The simulated event set imitates the recent PAO data \cite{Abreu2010arXiv10091855}, e.g., 69 events above $5.5 \times 10^{19}$ eV are distributed following the aperture geometry of the PAO. $B_{\rm EGMF} = 1$ nG and GMF are taken into account. The points and error bars of these cumulative auto-correlation functions are calculated as follows. First of all, 69 events are generated from a source distribution with the source number density of $n_s$, and then a cumulative auto-correlation function is calculated from this event set. Next, the former step is repeated over 1000 source distributions with the same $n_s$. Then, the averages and errors are calculated from the 1000 cumulative auto-correlation functions. When the number of events is small, the distribution of the values of the auto-correlation functions in a bin are asymmetric and therefore the standard deviation of the 1000 auto-correlation functions is not a good indicator of errors. In order to estimate anisotropic error bars, we arrange the values of the 1000 cumulative auto-correlation functions in each bin in order, calculating the range in which the central 68\% of the values are included, and then this range is regarded as a 68\% ($1\sigma$) error. This error includes not only an error due to the finite number of events (Poisson error) but also an error due to the uncertainty originating from the selection of UHECR sources. Finally, the averages and 68\% errors are plotted in the figure. All the figures in this paper follow this rule, as long as there is not exceptional description. 

\begin{figure}[t]
\begin{center}
\includegraphics[width=0.95\linewidth]{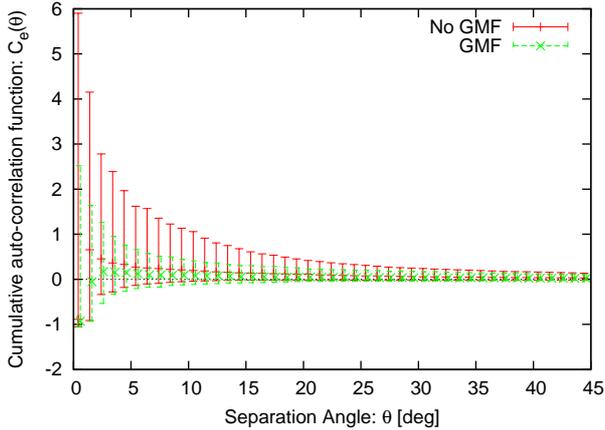}
\caption{Cumulative auto-correlation functions of 69 events above $5.5 \times 10^{19}$ eV on the assumption of the PAO aperture for $n_s = 10^{-4}$ Mpc$^{-3}$ and $B_{\rm EGMF} = 1$ nG. The GMF is considered ({\it green}) and is not considered ({\it red}) for comparison.}
\label{fig:GMF}
\end{center}
\end{figure}

All the auto-correlation functions except for $n_s = 10^{-7}$ Mpc$^{-3}$ are consistent with zero within 68\% errors in this angular range, i.e., the arrival distributions are consistent with isotropic distribution. Note that the case of $n_s = 10^{-6}$ Mpc$^{-3}$ corresponds to only 4 sources within nearby 100 Mpc. In spite of such a small number, the predicted arrival distribution is consistent with isotropy because of the strong effects of the GMF and EGMF. These magnetic fields make even the case of $n_s = 10^{-7}$ Mpc$^{-3}$ become consistent with isotropy within 95\% error at small angular scale ($\lesssim 20^{\circ}$).

The GMF significantly contributes to the arrival distribution of UHECRs. Fig. \ref{fig:GMF} shows the cumulative auto-correlation functions of 69 simulated events in the case of $n_s = 10^{-4}$ Mpc$^{-3}$ and $B_{\rm EGMF} = 1$ nG. In order to see the effect of the GMF, one ({\it green}) takes the GMF into account, but the other ({\it red}) does not consider the GMF. Only in this figure, the medians of 1000 cumulative auto-correlation functions are plotted instead of the averages for visibility because the very large values of cumulative auto-correlation functions are realized in some realizations by very nearby sources at small angular scale and the averaged values of the auto-correlation functions become larger than the upper edge of the 68\% error bars when the GMF is not taken into account. Note that the medians will be close to the averages when the number of events increases. Focusing on the first bin, we can see that the medians in both the cases are close to the lower edges of the error bars. This means that source distributions including nearby sources accidentally make strong anisotropy whereas no event pair appears in the first bin in many realizations because of the small solid angle within $1^{\circ}$ and the small number of simulated events. Thus, the distribution of the values of the cumulative correlation functions in the first bin has a long tail to large values. Although both the cumulative auto-correlation functions are consistent each other (and also with isotropic distribution) within 68\% errors, the error bars are suppressed when the GMF is considered, that is, the probability that large values of the auto-correlation function are realized becomes smaller. The GMF strongly weakens anisotropy produced by nearby sources.

\begin{figure}
\begin{center}
\includegraphics[width=0.95\linewidth]{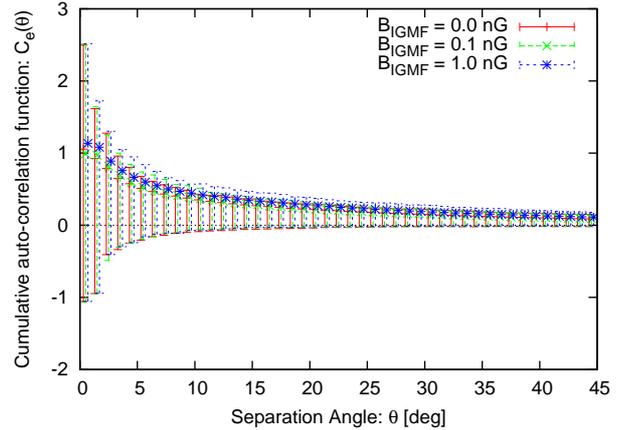}
\caption{Cumulative auto-correlation functions of 69 events above $5.5 \times 10^{19}$ eV simulated on the assumption of the PAO aperture for different $B_{\rm EGMF}$, $0.0$, $0.1$, and $1$ nG for comparison. $n_s = 10^{-5}$ Mpc$^{-3}$ and the GMF are considered.} 
\label{fig:GMFsup}
\end{center}
\end{figure}

Furthermore, we can find that the GMF can dominantly contribute to suppressing anisotropy. Fig. \ref{fig:GMFsup} shows the dependence of the cumulative auto-correlation functions on $B_{\rm EGMF}$ in the case of $n_s = 10^{-5}$ Mpc$^{-3}$ and considering the GMF. For comparison, $B_{\rm EGMF} = 0$, $0.1$, and $1$ nG are adopted. The three cumulative auto-correlation functions are quite similar, and therefore are controlled by the GMF. This situation is the same down to the case of $n_s = 10^{-6}$ Mpc$^{-3}$. Small difference of the averages among the cases of the three EGMF strengths appears only for $n_s = 10^{-7}$ Mpc$^{-3}$; $B_{\rm EGMF} \lesssim 0.1$ nG predicts a bit higher values of the averages, but the difference is still within the error bars. Thus, the effect of the GMF is essential and should be inevitably considered when the anisotropy of UHECR arrival distribution is tested.

\begin{figure*}
\begin{center}
\includegraphics[width=0.48\linewidth]{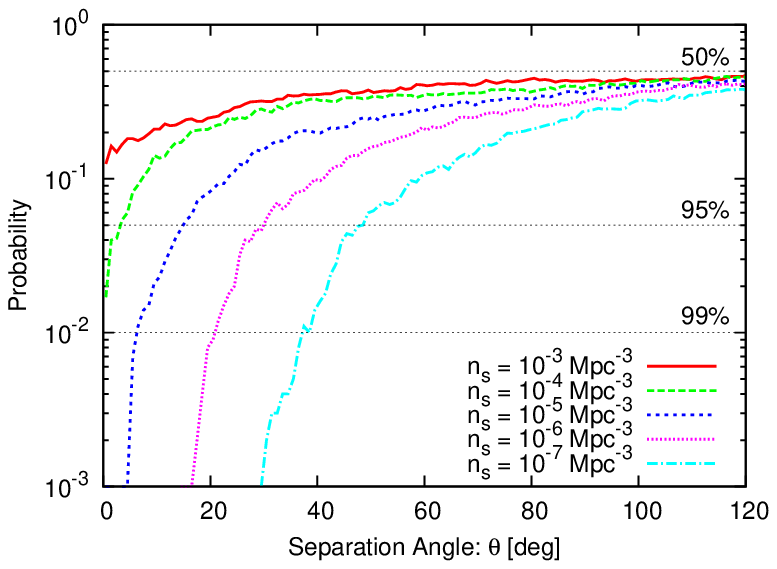} \hfill
\includegraphics[width=0.48\linewidth]{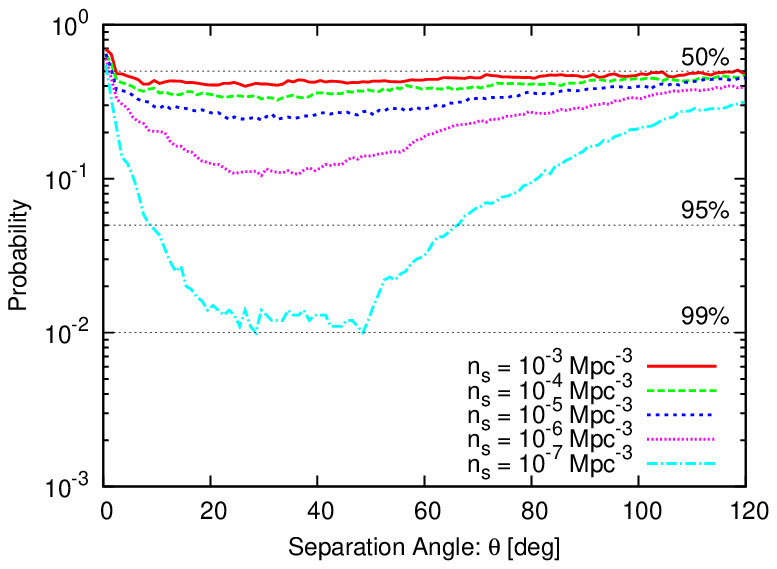} \\
\includegraphics[width=0.48\linewidth]{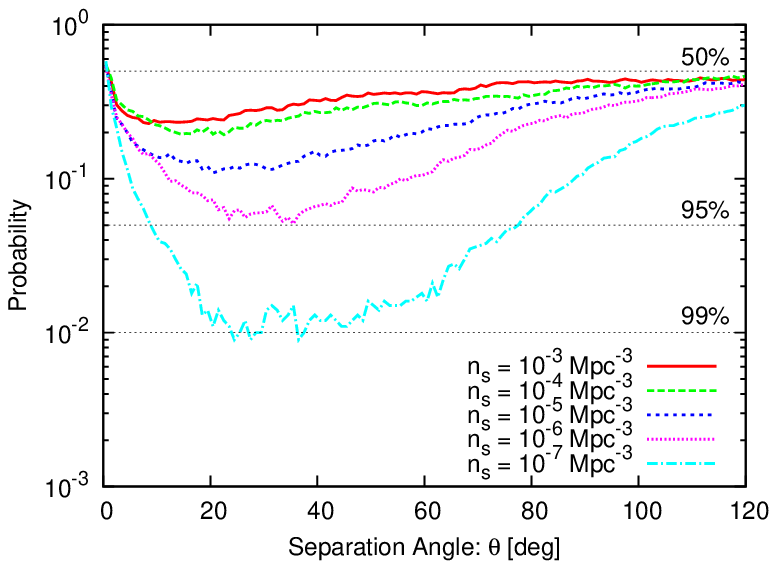} \hfill
\includegraphics[width=0.48\linewidth]{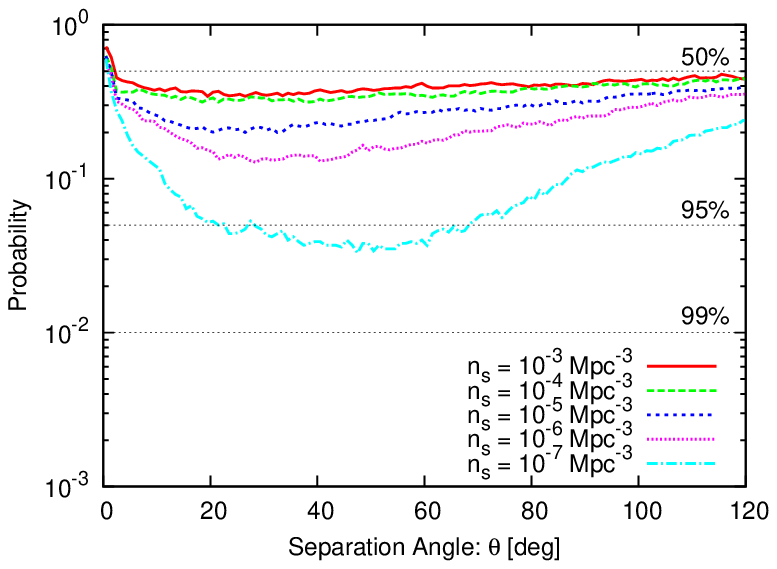}
\caption{Probability that the positive values of the cumulative auto-correlation functions are not realized for 69 events above $5.5 \times 10^{19}$ eV simulated on the assumption of the PAO aperture. The upper two panels and the lower two panels assume $B_{\rm EGMF} = 0.0$ nG and $1.0$ nG, respectively. Only the right two panels take the GMF into account. The probability that the positive excess of events appears is indicated in the panels. }
\label{fig:cprob_auto69}
\end{center}
\end{figure*}

Here, we are interested in how often a positive value of the auto-correlation function is realized in each bin. This provides the probability that the positive excess of events appears in each angular scale. This probability is calculated as follows. We simulate a set of UHECR events from a source distribution and calculate the cumulative auto-correlation function of them. This calculation is repeated over 1000 source distributions with the same $n_s$. Then, we focus on the values of the 1000 cumulative auto-correlation functions in an angular bin, counting the number of the auto-correlation functions which are positive, and the number is divided by 1000. This step is repeated over all the angular bins. Since only 1000 source distributions are considered (because of the computational limitation), it cannot resolve the probability less than $10^{-3}$, but it is enough for this study. By definition, isotropic distribution predicts this value of $\sim 50\%$ due to statistical fluctuation. This probability also can be regarded as a confidence level in the one-side statistical test to rule out isotropic distribution. For visibility, we plot one minus this probability throughout the paper, which is the probability that the positive excess of events is not realized.

Fig. \ref{fig:cprob_auto69} shows such probabilities for 69 simulated events above $5.5 \times 10^{19}$ eV. The aperture geometry of the PAO is applied. The upper two panels and the lower two panels are calculated under the conditions of $B_{\rm EGMF}=0$ nG and $1.0$ nG, respectively. The GMF is considered for only the right two panels.

In the case of no magnetic field ({\it upper left}), the strong positive excess of events appears at small angular scale. This is natural because the propagation trajectories are not deflected. In general, the probability that the positive excess of events is realized in a certain bin is higher for a smaller $n_s$ at small to intermediate angular scale because each source has stronger power and stronger clusterings of UHECR events are expected in the directions of their sources. The positive anisotropy is produced at more than 99\% for $n_s \lesssim 10^{-5}$ Mpc$^{-3}$ at small angular scale. Note that this does not mean that the positive excess signal of a set of observed events compared to isotropic distribution is larger than 99\% confidence level at the angular scale.

Magnetic fields change the shape of the probability curves. When the GMF is taken into account ({\it upper right}), the angular scale at which the value of a probability curve is minimized moves from the first angular bin to larger angular scale for all the $n_s$. The possibility of positive excess is maximized at this angular scale. This angular scale reflects the deflection angles of UHECR trajectories by the GMF. Since these are independent of the distances of sources, it does not depend on $n_s$. Although the EGMF gives a similar effect to the GMF ({\it lower left}), the angular scale of the minimum of a probability curve is different and depends on $n_s$ because the deflections of UHECRs by the EGMF are dependent on their propagation distance and the nearest sources, which contribute to anisotropy most strongly, are more distant in the case of a smaller $n_s$. In both cases only sources with $n_s = 10^{-7}$ Mpc$^{-3}$ produce positive excess in intermediate angular scale with more than 95 \% probability.

The result when both the GMF and EGMF are taken into account ({\it lower right}) can be interpreted as combination of the upper right and lower left panels. The contribution of the two magnetic fields reduces the probability that positive excess of events appears and moves the angular scale of the maximum of the probability to larger angular scale.

\begin{figure}
\begin{center}
\includegraphics[width=0.95\linewidth]{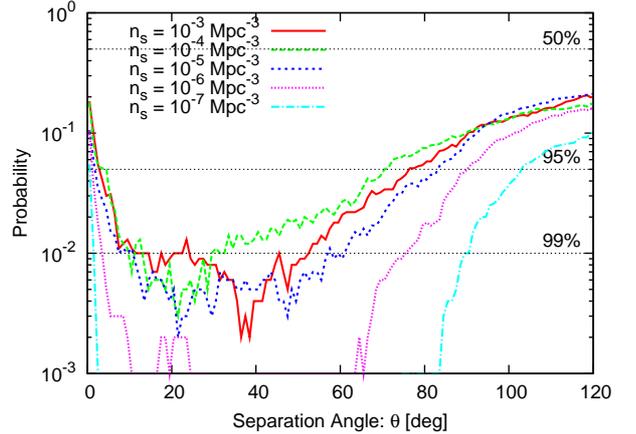}
\caption{Same as the lower right panel of Fig.\ref{fig:cprob_auto69}, but for 1000 events above $5.5 \times 10^{19}$ eV simulated on the assumption of a uniform aperture.} 
\label{fig:cprob_auto1000}
\end{center}
\end{figure}

At the end of this section we consider the case with the number of events expected by future UHECR experiments such as the Northern Pierre Auger Observatory \cite{Bluemer2010NJPh12c5001} and Extreme Universe Space Observatory on board Japanese Experiment Module (JEM-EUSO) \cite{JEMEUSO2008ICRC5p1045J}. Fig. \ref{fig:cprob_auto1000} is the same figure as that in the lower right panel of Fig. \ref{fig:cprob_auto69} (the GMF and EGMF are considered), but for 1000 events above $5.5 \times 10^{19}$ eV on the assumption of the uniform aperture. All the cases of $n_s$ predict the positive excess of events at intermediate angular scale at more than 99 \%. We notice that the case of $n_s = 10^{-3}$ Mpc$^{-3}$ shows a larger probability of the positive excess than that of $n_s = 10^{-4}$ Mpc$^{-4}$ at around $40^{\circ}$. The extremely close sources included for $n_s = 10^{-3}$ Mpc$^{-3}$ (within a few Mpc) causes this feature. When sources within 5 Mpc are artificially neglected for example to check this, the probability curves for $n_s \gtrsim 10^{-4}$ Mpc$^{-3}$ change significantly and the inversion disappears. In this case, the probability for $n_s = 10^{-3}$ Mpc$^{-3}$ becomes a bit smaller to 90\%.

\subsection{Correlation with sources} \label{ccor}

\begin{figure}
\begin{center}
\includegraphics[width=0.95\linewidth]{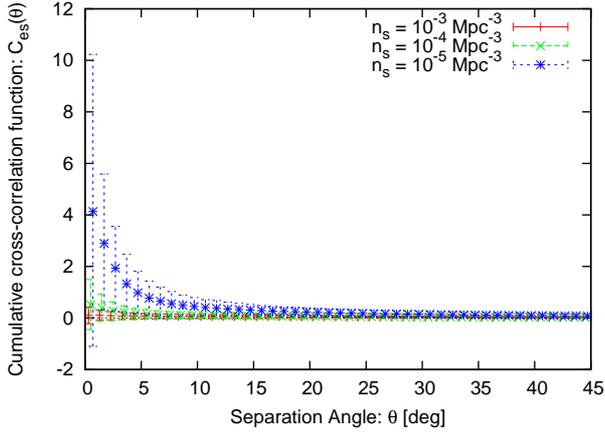}
\caption{Cumulative cross-correlation functions between the arrival directions of simulated UHECR events and the celestial positions of their sources with $n_s = 10^{-3}$, $10^{-4}$, and $10^{-5}$ Mpc$^{-3}$ within 75 Mpc. The simulated event sets imitate the published events of the PAO, i.e., 69 events with energies above $5.5 \times 10^{19}$ eV simulated on the assumption of the PAO aperture. Both EGMF ($B_{\rm EGMF} = 1$ nG) and the GMF are taken into account. The points are the averaged values of the cumulative cross-correlation functions over 1000 source realizations for each $n_s$. The error bars correspond to 68\% errors.} 
\label{fig:ccor1}
\end{center}
\end{figure}

Fig.\ref{fig:ccor1} shows the cumulative cross-correlation functions between the arrival directions of 69 UHECRs above $5.5 \times 10^{19}$ eV simulated on the assumption of the PAO aperture and the celestial positions of their sources within 75 Mpc with 68\% errors in the cases of $n_s = 10^{-3}$-$10^{-5}$ Mpc$^{-3}$. The 75 Mpc is motivated by the first report of the correlation between UHECRs and nearby extragalactic sources by the PAO \cite{Abraham2007Sci318p938}. $B_{\rm EGMF} = 1.0$ nG and the GMF are taken into account. The error bars and the averaged points of the cross-correlation functions are estimated similarly to those of cumulative auto-correlation functions. Note that we do not plot the cumulative cross-correlation functions for $n_s = 10^{-6}$ and $10^{-7}$ Mpc$^{-3}$ because there are source distributions not having sources within 75 Mpc in 1000 source realizations due to the small number densities. The cumulative cross-correlation functions for $n_s = 10^{-3}$ and $10^{-4}$ Mpc$^{-3}$ are fully consistent with zero within 68\% errors in this angular scale, i.e., are consistent with no correlation. The case of $n_s = 10^{-5}$ Mpc$^{-3}$ is positive at small angular scale with 68\% errors, but is consistent with no correlation within 95\% errors. The error bars are longer for smaller $n_s$ because the fluctuation of source positions is larger.

\begin{figure}
\begin{center}
\includegraphics[width=0.95\linewidth]{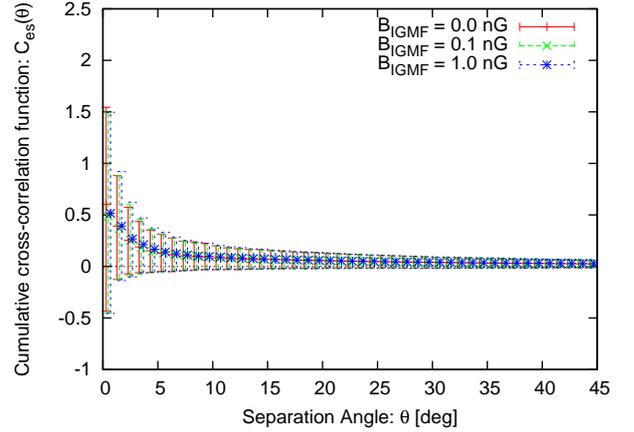}
\caption{Cumulative cross-correlation functions between 69 UHECRs above $5.5 \times 10^{19}$ eV simulated under the PAO aperture and sources within 75 Mpc for different $B_{\rm EGMF}$, $0.0$, $0.1$, and $1.0$ nG for comparison. $n_s = 10^{-4}$ Mpc$^{-3}$ and the GMF are considered.} 
\label{fig:ccor2}
\end{center}
\end{figure}

The cross-correlation between UHECRs and their sources is also dominantly affected by the GMF. Fig.\ref{fig:ccor2} is similar to Fig. \ref{fig:GMFsup}, but the cumulative cross-correlation functions between the 69 UHECRs and their sources with $n_s = 10^{-4}$ Mpc$^{-3}$ within 75 Mpc. $B_{\rm EGMF} = 0$ ({\it red}), $0.1$ ({\it green}), and $1$ nG ({\it blue}) are considered with the GMF for comparison. Similarly to Fig. \ref{fig:GMFsup}, the three cumulative cross-correlation functions are very similar. The situation is the same even for $n_s = 10^{-3}$ and $10^{-5}$ Mpc$^{-3}$. Even in this figure, we can notice the importance of the GMF on the discussions of the arrival distribution of UHECRs.

\begin{figure*}
\begin{center}
\includegraphics[width=0.48\linewidth]{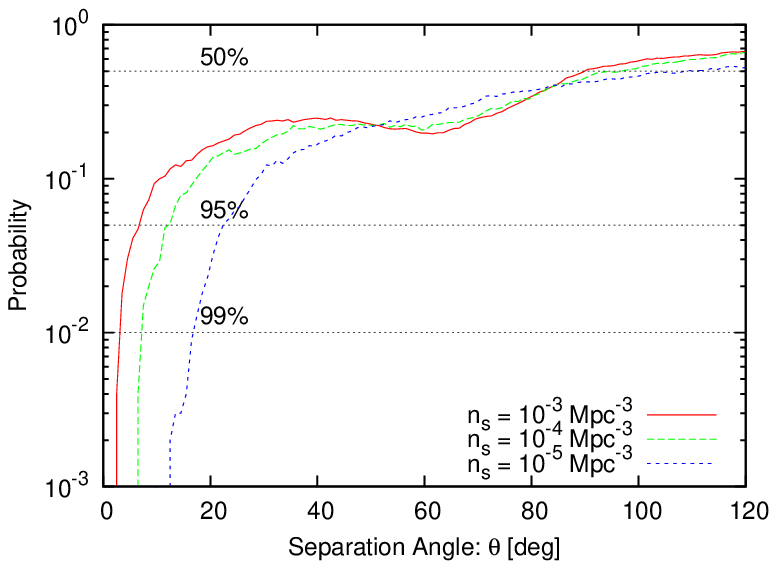} \hfill
\includegraphics[width=0.48\linewidth]{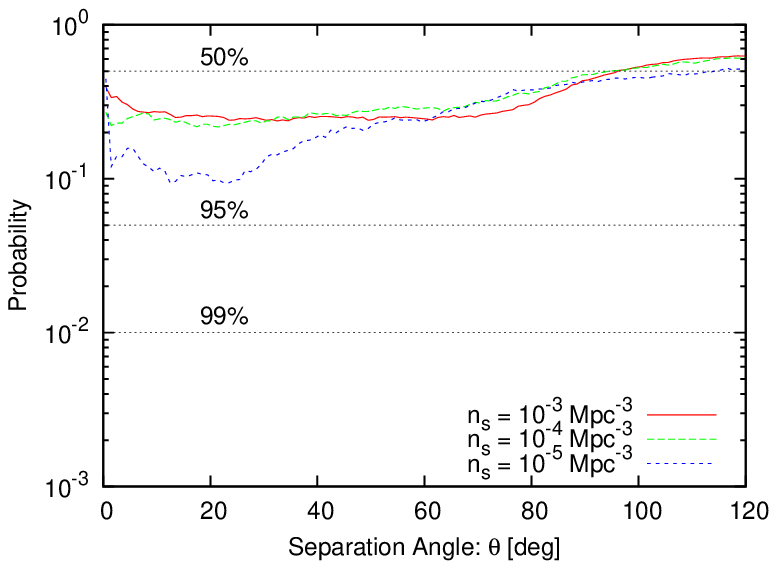} \\
\includegraphics[width=0.48\linewidth]{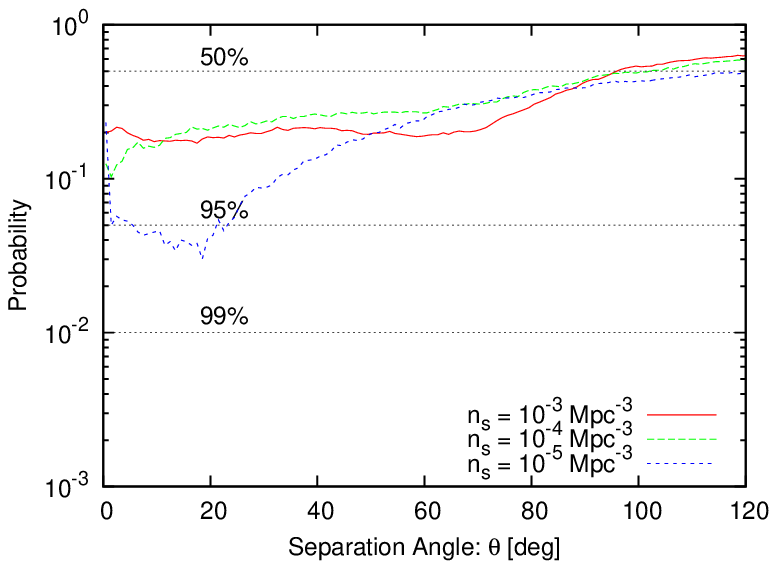} \hfill
\includegraphics[width=0.48\linewidth]{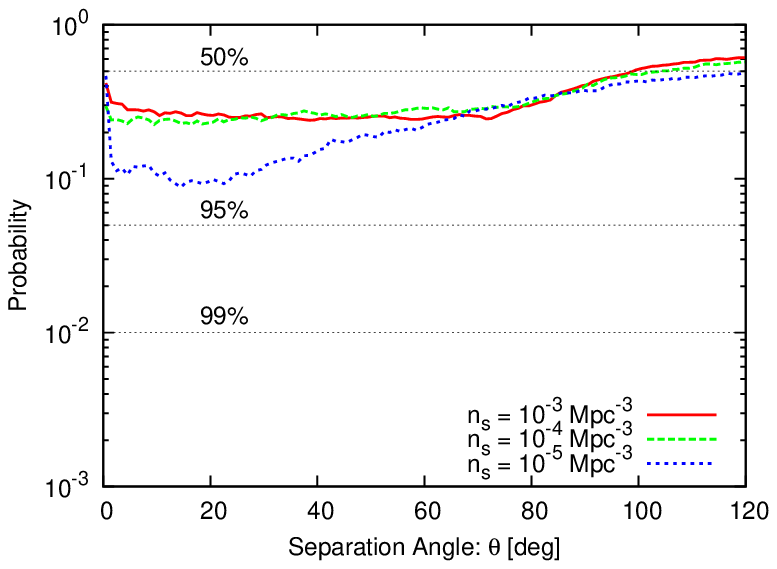}
\caption{Probability that the values of the cumulative cross-correlation function between the arrival directions of 69 UHECRs above $5.5 \times 10^{19}$ eV simulated under the PAO aperture and the celestial positions of their sources within 75 Mpc is not positive. The upper two panels and the lower two panels assume $B_{\rm EGMF} = 0.0$ nG and $1.0$ nG, respectively. Only the right two panels take the GMF into account. The probability that positive correlation appears is indicated in the panels.} 
\label{fig:ccorprob69}
\end{center}
\end{figure*}

Estimating the probability that positive correlation between UHECRs and their sources appears is also useful. We can calculate this probability by a similar method to calculate Fig. \ref{fig:cprob_auto69}. Following that figure, we plot the possibility that the values of the cross-correlation function are not positive for visibility in Fig.\ref{fig:ccorprob69} similarly to Fig. \ref{fig:cprob_auto69}. We consider 69 UHECRs above $5.5 \times 10^{19}$ eV on the assumption of the PAO aperture. The sources used for calculating the cross-correlation function are within 75 Mpc from the Galaxy. The basic tendency of the probability curves is similar to Fig.\ref{fig:cprob_auto69}. In the case of no magnetic field, strong correlation is predicted at the smallest angular bin because of the absence of the deflections of UHECRs. Taking the GMF and/or EGMF into account, the probability curves have a minimum at intermediate angular scale. For the cross-correlation function, the angular scale at which gives the minimum depends not only the deflection angles of UHECRs but also the typical separation angle between sources. The potential minimum is at $\sim 20^{\circ}$ for $n_s = 10^{-5}$ Mpc$^{-3}$ if the GMF and/or EGMF are considered, but the probability of the positive correlation is less than 95\%.

\begin{figure*}
\begin{center}
\includegraphics[width=0.48\linewidth]{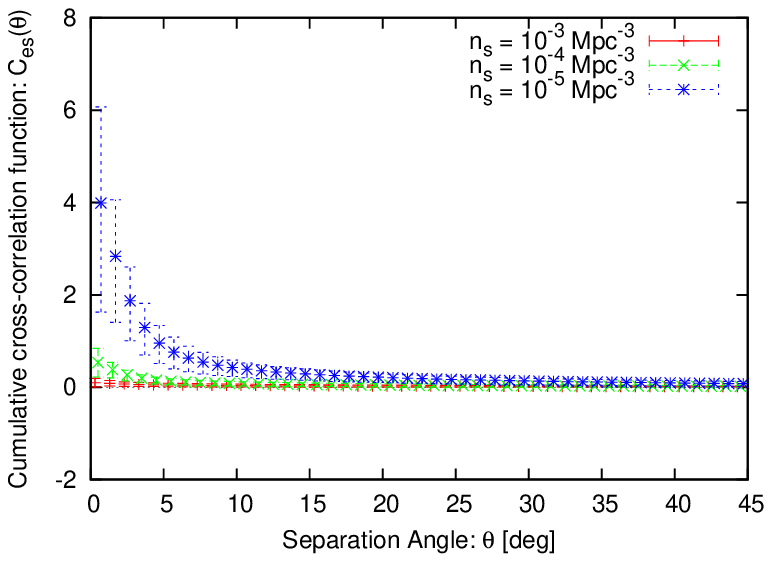}
\includegraphics[width=0.48\linewidth]{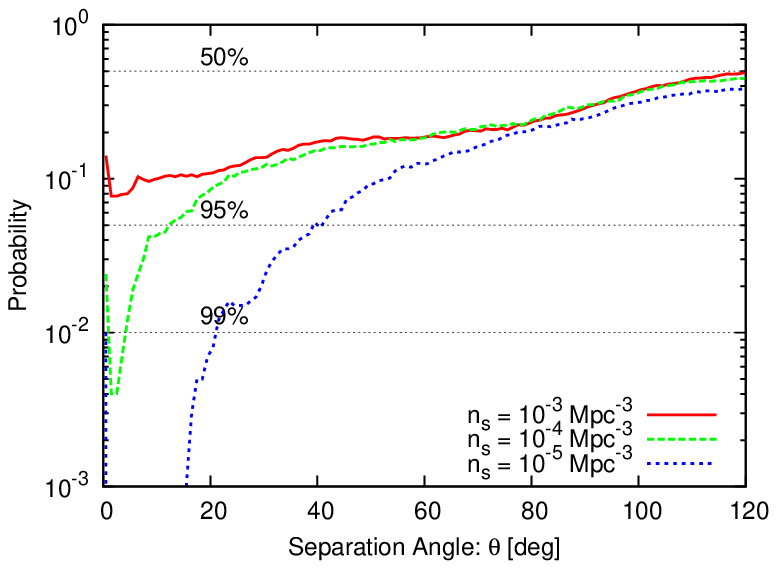}
\caption{({\it left}): Cumulative cross-correlation functions between 1000 UHECRs above $5.5 \times 10^{19}$ eV simulated on the assumption of a uniform aperture and the celestial positions of their sources within 75 Mpc. EGMF ($B_{\rm EGMF} = 1.0$ nG) and the GMF are taken into account. ({\it right}): Probability that the positive values of the cumulative cross-correlation functions shown are not realized. The probability that positive correlation appears is indicated in the panel.}
\label{fig:ccorprob1000}
\end{center}
\end{figure*}

Again, we demonstrate the prospects of future UHECR observatories. Fig.\ref{fig:ccorprob1000} shows cumulative cross-correlation functions ({\it left}) and the corresponding probability curves ({\it right}). We consider 1000 UHECRs above $5.5 \times 10^{19}$ eV on the assumption of the uniform aperture and both the GMF and EGMF ($B_{\rm EGMF} = 1$ nG). In the left panel, the cumulative cross-correlation functions in the cases of $n_s = 10^{-4}$ and $10^{-5}$ Mpc$^{-3}$ are clearly separated from zero within 68\% errors, in comparison to Fig. \ref{fig:ccor1}. Although it is difficult to see, even the cases of $n_s = 10^{-3}$ Mpc$^{-3}$ is also inconsistent with no correlation within 68\% errors. The probability that the positive correlation appears is shown in the right panel. The cases of $n_s = 10^{-4}$ and $10^{-5}$ Mpc$^{-3}$ produce the correlation with more than 99 \% at small angular scale. Since stronger correlation is expected for the smaller $n_s$, next generation UHECR experiments can observe positive correlation between UHECRs and their sources at small angular scale with more than 99 \% probability if $n_s \lesssim 10^{-4}$ Mpc$^{-3}$.

\begin{figure*}[t]
\begin{center}
\includegraphics[width=0.95\linewidth]{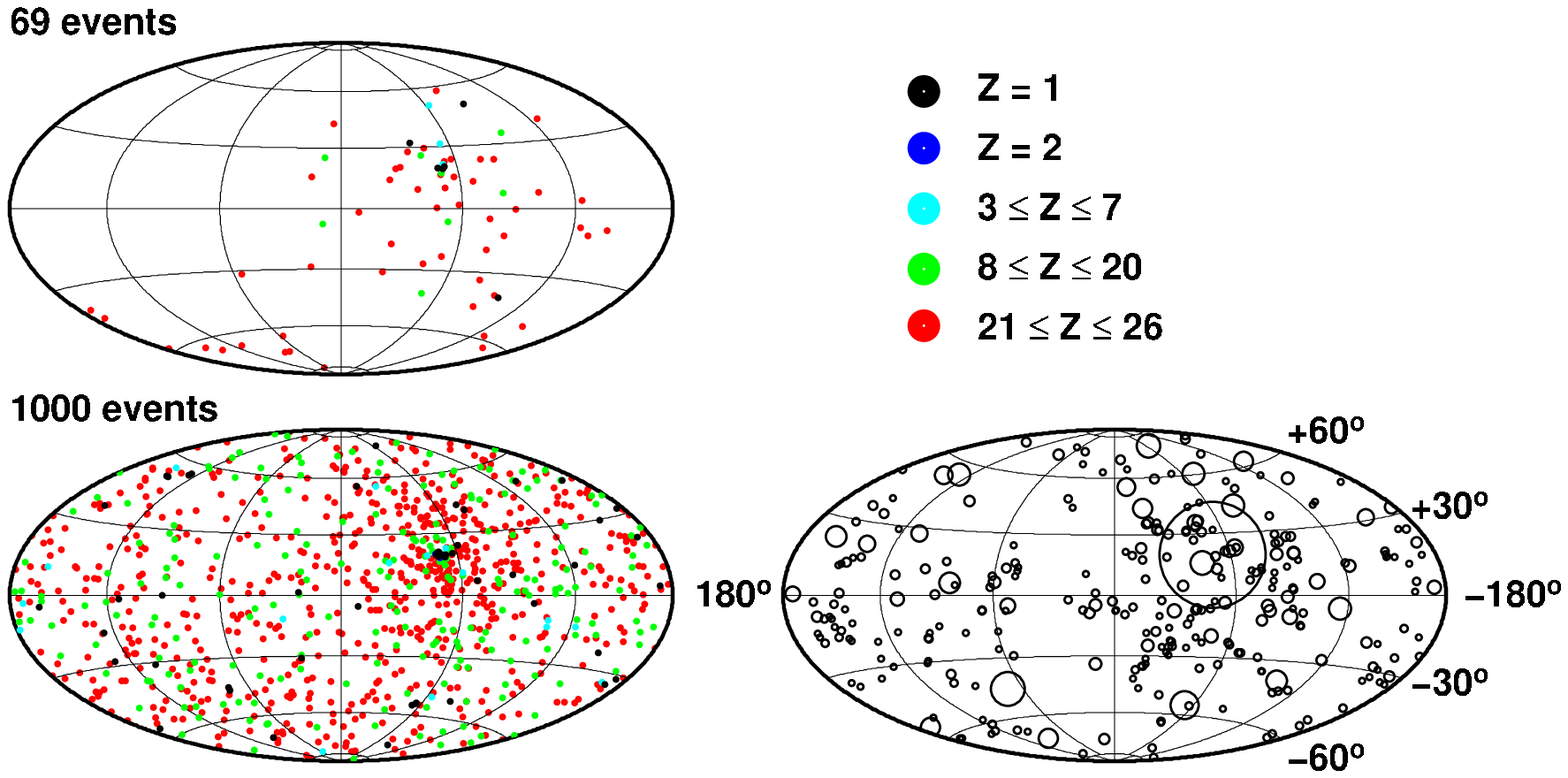}
\caption{Arrival distribution of UHECRs realized from a specific source distribution including Cen A with $n_s = 10^{-4}$ Mpc$^{-3}$. 69 events above $5.5 \times 10^{19}$ eV under the PAO aperture ({\it upper left}) and 1000 events above $5.5 \times 10^{19}$ eV under a uniform aperture ({\it lower right}) are considered. $E_{\rm max}^{\rm Fe} = 10^{21.5}$ eV is assumed. The composition of the UHECRs is represented in color. The source distribution used in this calculation is shown in the lower right panel, but only the sources within 75 Mpc are plotted as circles for visibility. The radii of the circles are inversely proportional to the distances of sources. The position of Cen A is the center of the largest circle.}
\label{fig:map}
\end{center}
\end{figure*}

\subsection{Cen A as Nearest UHECR Source} \label{cena}

The recent PAO data with energies above $5.5 \times 10^{19}$ eV show the overdensity of the arrival directions in the direction of Cen A and that 2 events positionally correlate with the nucleus position of Cen A \cite{Abreu2010arXiv10091855}. On the other hand, Cen A has been expected as nearby production site of UHECRs \cite{Cavallo1978AA65p415,Romero1996APh5p279,Fraschetti2008MNRAS391p1100,Hardcastle2009MNRAS393p1041,OSullivan2009MNRAS400p248O}. In this section, we specially focus on Cen A, and discuss the possibility that Cen A is the nearest UHECR source from the viewpoint of the arrival distribution of UHECRs.

For this purpose, we use two source models which are slightly modified versions of the source model introduced in Section \ref{method_ad}. In the first model, we artificially neglect sources within 5 Mpc from the original source model and add Cen A. In this source model, all the source distributions include Cen A as the nearest UHECR source. We assume that the positions and distance of Cen A are ($\ell$, $b$) $=$ ($319^{\circ}$, $19.5^{\circ}$) and 4.1 Mpc following parameters listed in the IRAS PSCz catalog \cite{Saunders2000MNRAS317p55}. The second source model is for reference, i.e., we artificially neglect sources within 5 Mpc from the original model but do not add Cen A. Comparing the results from these two source models, we can investigate the effect of Cen A to UHECR arrival distribution.

Fig. \ref{fig:map} demonstrates the arrival distributions of UHECRs above $5.5 \times 10^{19}$ eV from a specific source distribution including Cen A with $n_s = 10^{-4}$ Mpc$^{-3}$ ({\it lower right}). This number density is comparable with the local number density of FR I galaxies, which Cen A is classified into \cite{Padovani1990ApJ356p75}. The source distribution is shown as circles in the Galactic coordinates. The radii of the circles are inversely proportional to the distances of sources. Although we use all the sources up to 500 Mpc in the calculations, only sources within 75 Mpc are plotted as circles for visibility. The position of Cen A can be identified as the center of the largest circle. The simulated events are 69 under the PAO aperture ({\it upper left}) and 1000 under the uniform aperture ({\it lower left}). The colors represent the composition of UHECRs; protons ({\it black}), helium ({\it blue}), light nuclei with $3 \leq Z \leq 7$ ({\it light blue}), intermediate mass nuclei with $8 \leq Z \leq 20$ ({\it green}), and heavy nuclei with $21 \leq Z \leq 26$ ({\it red}), where $Z$ is the nuclear number.

Even the 69 events remind you that a significant fraction of the whole events is distributed around Cen A. In this case, about 30\% of UHECRs originates from Cen A. Nevertheless, the arrival directions of the UHECRs are scattered by GMF (see also Ref. \cite{Giacinti2011APh35p192}). Several black points are secondary protons. Since protons are minimally affected by the GMF, they can point out their sources. Three protons correlate with the position of Cen A very well. If Cen A is a true source, it is possible that the two PAO events correlating with the position of Cen A are protons, because several plausible GMF models predict very small deflection angles ($\sim 1^{\circ}$) in the direction of Cen A \cite{Takami:2007kq}. Note that secondary protons above $5.5 \times 10^{19}$ eV require the maximum acceleration energy of irons at sources above $10^{21.5}$ eV.

\begin{figure}[t]
\begin{center}
\includegraphics[width=0.95\linewidth]{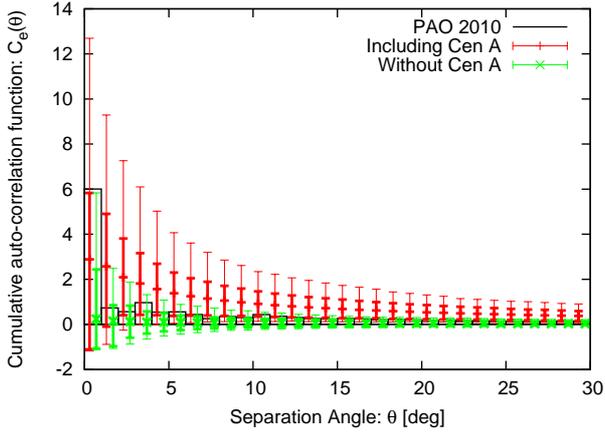} \hfill
\caption{Cumulative auto-correlation functions of the arrival directions of 69 UHECRs with the energy above $5.5 \times 10^{19}$ eV simulated under the PAO aperture. The source models including ({\it red}) and neglecting ({\it green}) Cen A with $n_s = 10^{-4}$ Mpc$^{-3}$ are used. Both EGMF ($B_{\rm EGMF} = 1$ nG) and GMF are taken into account. The error bars represent 68\% ({\it thick}) and 95\% ({\it thin}) errors. The cumulative auto-correlation function calculated from the recent published PAO data \cite{Abreu2010arXiv10091855} is also shown for comparison. ({\it black}).} 
\label{fig:CenAornot1}
\end{center}
\end{figure}

The reproducibility of the auto-correlation function of the observed data by the source model including Cen A should be checked. Fig. \ref{fig:CenAornot1} shows the cumulative auto-correlation functions of the arrival directions of 69 UHECRs above $5.5 \times 10^{19}$ eV simulated on the assumption of the PAO aperture. The source models with $n_s = 10^{-4}$ Mpc$^{-3}$ explained above are applied. The error bars represent 68\% ({\it thick}) and 95\% ({\it thin}) errors. Both the GMF and EGMF ($B_{\rm EGMF} = 1$ nG) are taken into account. The cumulative auto-correlation function of the PAO events is also shown for reference. This figure shows that the predictions of both source models are consistent with the recent PAO data within 95\% errors, even in the case that Cen A is the nearest UHECR sources. On the other hand, the arrival distribution of the recent PAO data is also consistent with isotropic distribution above $5.5 \times 10^{19}$ eV at 95\% confidence level \cite{Abreu2010APh34p314}. Thus, both possibilities, Cen A is a source or isotropic, are allowed at present. Note that the meaning of the 95\% in the PAO analysis is different from that of the 95\% in this analysis. The observed arrival distribution is consistent with the predictions of our source model at 95\% confidence level in this study, while the distribution is consistent with isotropic distribution at 95\% confidence level in the analysis of the PAO.

\begin{figure*}[t]
\begin{center}
\includegraphics[width=0.48\linewidth]{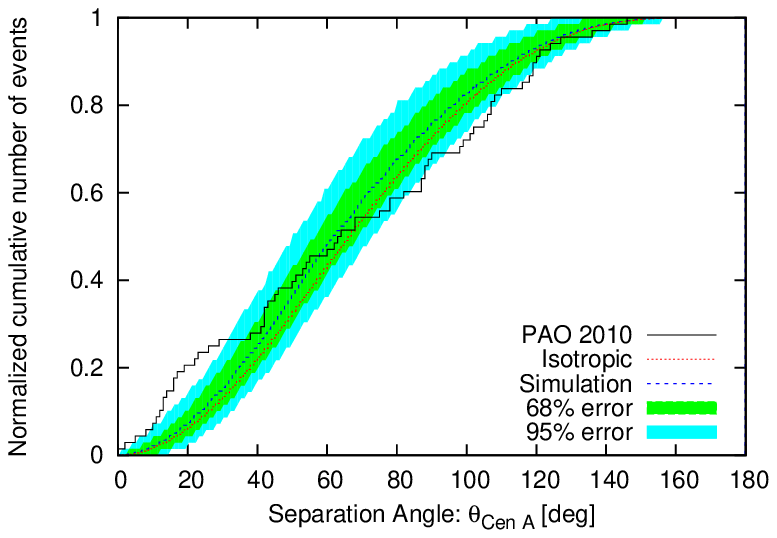} \hfill
\includegraphics[width=0.48\linewidth]{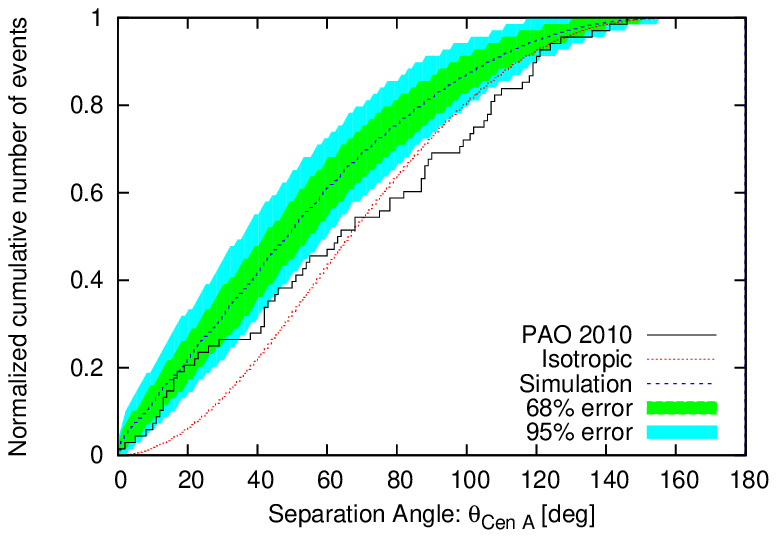}
\caption{Normalized cumulative numbers of 69 events with energies above $5.5 \times 10^{19}$ eV simulated under the PAO aperture as a function of angular distance from the nucleus position of Cen A. In simulations, source models with $n_s = 10^{-4}$ Mpc$^{-3}$ including Cen A ({\it right}) and neglecting Cen A ({\it left}) are used. Both EGMF ($B_{\rm EGMF} = 1$ nG) and the GMF are taken into account. The blue lines are the averages of the cumulative numbers of the simulated events. The green and light blue regions represent 68\% and 95\% errors, respectively. For comparison, the event distribution of the 69 PAO events listed in \cite{Abreu2010arXiv10091855} ({\it black}) and that of isotropic events ({\it red}) are also shown.} 
\label{fig:cenadist}
\end{center}
\end{figure*}

The PAO data also reported significantly positive excess of UHECRs in the arrival distribution around the nucleus position of Cen A \cite{Abreu2010arXiv10091855}. This excess gives information to test whether Cen A significantly contributes to the total UHECR flux or not. Fig. \ref{fig:cenadist} shows the normalized cumulative numbers of events used in Fig. \ref{fig:CenAornot1}. The average of the cumulative numbers is represented as {\it blue} lines with 68\% ({\it green}) and 95\% ({\it light blue}) errors. The event distribution of the 69 PAO events ({\it black}) and that of isotropic events ({\it red}) are also shown for comparison.

The source model without Cen A can well reproduce the event distribution of the PAO events above $30^{\circ}$ within 95\% error bars ({\it left}). The prediction is also consistent with the distribution of isotropic events. However, this source model cannot reproduce the excess around Cen A despite that our source model includes the overdensity of the Centaurus region. On the other hand, the source model including Cen A reproduces the excess around Cen A ({\it right}). Although the predicted distribution cannot reproduce the observed event distribution at large angular scales, it indicates that the contribution of Cen A is important to reproduce this excess.

In order to reproduce this observed distribution over all the angular scale, the contribution of Cen A has to be reduced or the deficit of events should exist at large angular scale from Cen A. Possible solutions of the former idea are (i) UHECR flux of Cen A is weaker than that of the other FR I galaxies, (ii) there is an additional source population with $n_s \gtrsim 10^{-4}$ Mpc$^{-3}$ like Seyfert galaxies \cite{Pe'er2009PRD80p123018} and the relative contribution of Cen A is reduced, (iii) GMF can work more efficiently than what we consider in this study. For the latter idea a possibility is Cen A mainly contributes to UHECRs in the whole sky (at least southern sky) but the UHECRs are not completely isotropized with keeping consistency with the observed isotropy mentioned above. In any case, scans over large parameter space are needed, which is beyond the scope of this paper. Here, we only claim that the absence of Cen A cannot reproduce the observed excess.

The increasing number of events reveals anisotropy by nearby sources more clearly, as seen in the lower left panel of Fig. \ref{fig:map}. If Cen A significantly contributes to the total flux of UHECRs, anisotropy in the arrival distribution at angular scale of $\lesssim 30^{\circ}$ clearly appears and the order of tens secondary protons correlates with the nucleus position of Cen A. In this case, $\sim 16\%$ of the events originates from Cen A. In addition, a significant fraction of secondary protons positionally correlates with nearby sources other than Cen A. These secondary protons are an indicator of nearby sources even if only irons are accelerated at sources.

\begin{figure*}[t]
\begin{center}
\includegraphics[width=0.48\linewidth]{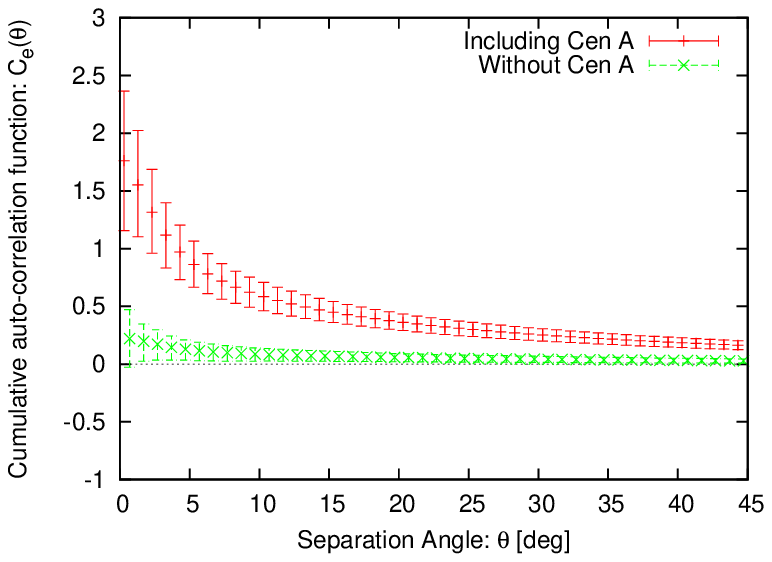} \hfill
\includegraphics[width=0.48\linewidth]{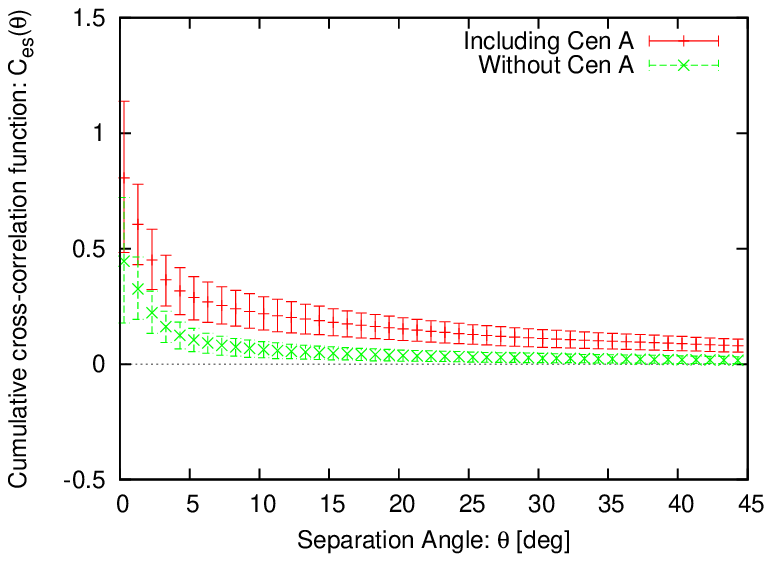}
\caption{({\it left}): Cumulative auto-correlation functions of the arrival directions of 1000 simulated events above $5.5 \times 10^{19}$ eV. ({\it right}): Cumulative cross-correlation functions between 1000 simulated events above $5.5 \times 10^{19}$ eV and their sources within 75 Mpc. In both panels, a uniform aperture, EGMF ($B_{\rm EGMF} = 1$ nG) and the GMF are assumed. The error bars represent 68\% errors.}
\label{fig:CenAornot2}
\end{center}
\end{figure*}

Fig. \ref{fig:CenAornot2} shows the cumulative auto-correlation functions ({\it left}) and cumulative cross-correlation functions ({\it right}) of 1000 events above $5.5 \times 10^{19}$ eV simulated on the assumption of the uniform aperture. The source models including Cen A ({\it red}) and neglecting Cen A ({\it green}) with $n_s = 10^{-4}$ Mpc $^{-3}$ are applied, and both EGMF ($B_{\rm EGMF} = 1$ nG) and GMF are taken into account. The two source models are well separated in the cumulative auto-correlation functions, i.e., anisotropy in events detected in the future will be a good clue to understand the contribution of nearby UHECR sources. On the other hand, although the two source models are not separated in the cumulative cross-correlation functions as good as in the auto-correlation functions, both the cases predict significantly positive correlation between UHECRs and their sources at small angular scale, clearly. Even if primary UHECRs are purely irons, information on the location of nearby sources can be extracted from the future UHECR data.

\section{Discussion \& Summary} \label{summary}

\begin{figure}[t]
\begin{center}
\includegraphics[width=0.95\linewidth]{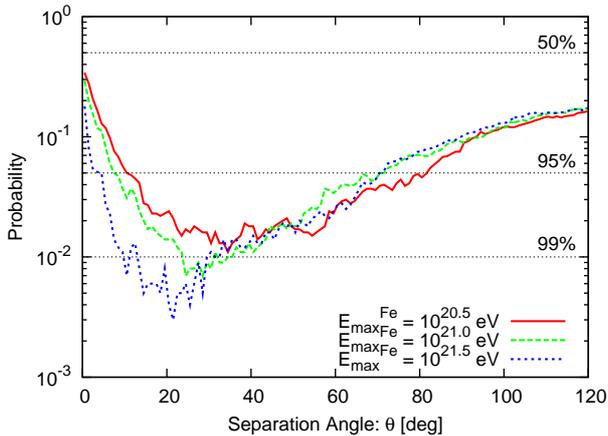} \hfill
\caption{Same as Fig. \ref{fig:cprob_auto1000}, but considering different 
$E_{\rm max}^{\rm Fe}$ indicated in the figure and $n_s = 10^{-4}$ Mpc$^{-3}$.}
\label{fig:acorEmax}
\end{center}
\end{figure}

We have investigated the anisotropy in the arrival directions of UHECR nuclei and their positional correlations with source objects, on the basis of propagation calculations conducted under various assumptions for the source properties and cosmic magnetic fields. Although $E_{\rm max}^{\rm Fe} = 10^{21.5}$ eV was fiducially assumed in the previous section, the results depend somewhat on $E_{\rm max}^{\rm Fe}$. Since only iron nuclei above $10^{21.5}$ eV can produce secondary protons above $5.5 \times 10^{19}$ eV ( $E_{\rm max}^{\rm p} = E_{\rm max}^{\rm Fe} / 56 = 5.6 \times 10^{19}$ [ $E_{\rm max}^{\rm Fe} / 10^{21.5}$ eV ] eV ), the threshold energy for the published PAO events and also adopted in this study, secondary protons do not contribute to the total flux of UHECRs above it if $E_{\rm max}^{\rm Fe} < 10^{21.5}$ eV. Fig. \ref{fig:acorEmax} shows the probability curves as discussed in Fig. \ref{fig:cprob_auto1000}, but for $E_{\rm max}^{\rm Fe} = 10^{20.5}$, $10^{21.0}$, and $10^{21.5}$ eV, in order to check the dependence on $E_{\rm max}^{\rm Fe}$. The probability for the appearance of a positive excess of events is the highest for $E_{\rm max}^{\rm Fe} = 10^{21.5}$ eV at small angular scales, and the angular scale of the peak is smaller for higher $E_{\rm max}^{\rm Fe}$. These reflect the fact that secondary protons contribute only in the case of $E_{\rm max}^{\rm Fe} = 10^{21.5}$ eV. Furthermore, the probability is slightly higher at small angular scales for $E_{\rm max}^{\rm Fe} = 10^{21.0}$ eV than for $E_{\rm max}^{\rm Fe} = 10^{20.5}$ eV because light and/or high energy nuclei are included above the threshold in the former case. Therefore, positive anisotropy at small angular scales is weakened when $E_{\rm max}^{\rm Fe} < 10^{21.5}$ eV, owing to the lack of secondary protons above threshold, but the probability at intermediate angular scales is still greater than 95\%.

We have considered a rather extreme case where only iron nuclei are emitted from sources. However, other nuclei and/or protons may also plausibly constitute the composition of UHECR sources (e.g., a mixed composition model \cite{Allard:2005cx} or a Wolf-Rayet star model \cite{Liu2012ApJ746p40}). The other elements can affect the spectral shape as mentioned in Section \ref{introduction} due to different energy losses and magnetic deflections. Although the source composition is highly uncertain at the moment, further studies of anisotropy and correlations under more realistic assumptions for the source composition are in order. For instance, assuming a mixed composition and a simple rigidity scaling for the maximum energies, protons can produce anisotropy at low energies comparable to or stronger than the anisotropy due to heavy nuclei at the highest energies, because a proton with energies $Z$ times smaller than an iron nucleus propagates along the same trajectory \cite{Lemoine2009JCAP11p009}. Thus, if Cen A is an UHECR source, anisotropy at low energies may also be expected. However, PAO has reported the lack of significant anisotropy at low energies toward the direction of the excess highest energy events  \cite{Abreu2011JCAP06p022}. This result implies that 1) a significant fraction of UHECRs may be nuclei even at lower energies, 2) the anisotropy at the highest energies is actually weaker than reported, 3) the composition at high energies is not as heavy nuclei-dominant as indicated from the X{\rm max} measurements, or some combination of these three possibilities. Furthermore, strong magnetic fields could modify the theoretical expectation through the extension of the propagation distance as mentioned in \cite{Abreu2011JCAP06p022}, so magnetic fields may also be important in this context.

Although we have adopted simple models for the GMF and EGMF to avoid the uncertainties with their structure, the details of these fields can actually play important roles in the propagation of UHECRs and their arrival directions at Earth, as has been well studied for UHE protons \cite{Sigl2003PRD68p043002,Sigl2004PRD70p043007,Dolag2005JCAP01p009,Takami2006ApJ639p803,Das2008ApJ682p29,Kotera2008PRD77p123003}. In particular, as discussed in Ref. \cite{Dolag2005JCAP01p009,Takami2006ApJ639p803,Kotera2008PRD77p123003} for protons, the structures actually observed near the Galaxy are also relevant when searching for the evidence of nearby UHECR sources. Whether the positional information in the sky is lost or not during propagation depends on the EGMF strength and structures in local Universe. The effect of the GMF is inevitable, but the deflection angles and directions of UHECRs strongly depend on the region of the sky, reflecting the structure of the GMF, as studied in Refs. \cite{Giacinti2010JCAP08p036,Giacinti2011APh35p192} for iron nuclei. Thus more realistic GMF models should be applied for correlation studies, although such GMF models also have significant uncertainties. The dependence of the anisotropy signals on different models for the GMF and EGMF will be investigated in future work.

In addition to Cen A, M87 is also an interesting object as an UHECR source candidate. Since it is more distant than Cen A ($\sim 16$ Mpc), the contribution of M87 to the anisotropy is smaller, depending on the EGMF strength. If the power of UHECR injection from M87 is comparable to that of typical FR I galaxies, it is not expected to contribute strongly to the anisotropy. However, if its power is an order of magnitude stronger or $n_s$ is smaller, anisotropy can appear in the direction of M87 even for $B_{\rm EGMF} = 1$ nG. Secondary protons should also be a good source indicator in this case.

This study has assumed the sources of UHECRs to be steady. On the other hand, it is also very much possible that transient phenomena like GRBs or AGN flares produce UHECRs. Transient scenarios are especially relevant for accelerating protons up to $10^{20}$ eV in outflows because not many steady sources satisfy the luminosity requirement for UHECR acceleration \cite{Norman1995ApJ454p60,Waxman:2003uj}. For transient sources, the arrival distribution of UHECRs will be different from that in the case of steady source scenarios. The time delay of UHECRs by the GMF and EGMFs leads to apparent durations of UHECR bursts longer than their intrinsic duration. If the apparent duration is longer than the observation timescale of human beings, we would misperceive an UHECR burst as a steady source \cite{MiraldaEscude1996ApJ462L59,Murase2008ApJ690L14,Takami2011arXiv1110.3245}. However, since the arrival time of UHECRs depends on the UHECR energies, features of anisotropy may be strongly energy-dependent \cite{MiraldaEscude1996ApJ462L59,Takami2011arXiv1110.3245}. This will be evidence for the transient generation of UHECRs. Even in transient scenarios, anisotropy of (secondary) protons at small angular scale should still be a strong hint of UHECR sources.

To summarize, we have calculated the propagation of UHE nuclei above $5.5 \times 10^{19}$ eV, taking into account magnetic fields in the universe on the assumption that only iron nuclei are injected by steady sources, and examined the anisotropy in the distribution of UHECR arrival directions and positional correlation between UHECRs and their sources in local universe. We found that both the anisotropy and correlation are dominantly suppressed by the GMF rather than by EGMFs. Assuming the current PAO status (69 events above $5.5 \times 10^{19}$ eV), the arrival distribution of UHECRs is consistent with isotropy within 95\% errors for $n_s \gtrsim 10^{-6}$ Mpc$^{-3}$ and is consistent with no correlation within 95\% errors for $n_s \gtrsim 10^{-5}$ Mpc$^{-3}$ when the GMF and EGMF are taken into account. We also discussed whether future experiments improve these situation. 1000 events above $5.5 \times 10^{19}$ eV in the whole sky would reveal positive anisotropy with more than 99\% probability at intermediate angular scales even for $n_s \sim 10^{-3}$ Mpc$^{-3}$, and correlation between UHECRs and their sources in local universe would appear with more than 99\% at small angular scales for $n_s \lesssim 10^{-4}$ Mpc$^{-3}$. In addition, we found that the contribution of Cen A is required to reproduce the observed UHECR excess around the Centaurus region. Secondary protons from primary heavy nuclei are working positively to produce anisotropy at small angular scales, which would provide a strong hint of the source location, depending on the maximum energy at the UHECR source. 

\subsubsection*{Acknowledgements:} 
We thank D.~Allard, G.~Medina-tanco, and F.~Takahara for useful comments. This work is supported by Grant-in-Aid No.~22540278 (SI) from the Ministry of Education, Culture, Sports, Science and Technology of Japan.
\bibliographystyle{model1a-num-names}

\end{document}